\documentclass[aps,twocolumn]{revtex4}
\usepackage{epsfig}
\newcommand{\be}{\begin{equation}}
\newcommand{\ee}{\end{equation}}
\newcommand{\bea}{\begin{eqnarray}}
\newcommand{\eea}{\end{eqnarray}}
\newcommand{\ba}{\begin{array}}
\newcommand{\ea}{\end{array}}

\begin{document}
\title{Least Dependent Component Analysis Based on Mutual Information}
\author{Harald St\"ogbauer, Alexander Kraskov, Sergey A. Astakhov, and Peter Grassberger}
\affiliation{John-von-Neumann Institute for Computing, Forschungszentrum
J\"ulich, D-52425 J\"ulich, Germany}

\date{\today}

\begin{abstract}
We propose to use precise estimators of mutual information (MI) to find least dependent 
components in a linearly mixed signal. On the one hand this seems to lead to better blind 
source separation than with any other presently available algorithm. On the other hand it 
has the advantage, compared to other implementations of `independent' 
component analysis (ICA) some of which are based on crude approximations for MI,
that the numerical values of the MI can be used for:\\
(i) estimating residual dependencies between the output components; \\
(ii) estimating the reliability of the output, by comparing the pairwise MIs with those 
of re-mixed components;\\
(iii) clustering the output according to the residual interdependencies.\\
For the MI estimator we use a recently proposed $k$-nearest neighbor based algorithm.
For time sequences we combine this with delay embedding, in order to take into account 
non-trivial time correlations. After several tests with artificial data, we apply the 
resulting MILCA (Mutual Information based Least dependent Component Analysis) algorithm
to a real-world dataset, the ECG of a pregnant woman.

\end{abstract}
\maketitle

\section{Introduction}

`Independent' component analysis (ICA) is a statistical method for transforming an observed 
multi-component data set ${\bf x}(t)=(x_1(t),x_2(t),...,x_n(t))$ into components that are 
statistically as independent from each other as possible \cite{hyvar2001}. In theoretical
analyses one usually assumes a certain model for the data for which a decomposition into 
completely independent components is possible, but in real life applications the latter 
will in general not be true. Depending on the assumed structure of the data, one 
typically makes a parametrized guess about how they can be decomposed (linearly or not, 
using only equal times or using also delayed superpositions, etc.) and then fixes the 
parameters by minimizing some similarity measure between the output components.

Using mutual information (MI) would be the most natural way to solve this problem. But
estimating MI from statistical samples is not easy. Most existing algorithms are either 
very slow or very crude. Also, the more sophisticated estimates usually do not depend
smoothly on transformations of the data, which slows down minimum searches. In the ICA 
literature mostly very crude approximations of MI are used, or MI is completely disregarded
in favor of different approaches \cite{hyvar2001,Amaribook}. In particular, we are aware of only
very few attempts to pay attention to the actual values of the similarities / (in)dependences
obtained by ICA. Of course it has been recognized several times that even the best decomposition 
with a given class of algorithms (e.g. linear and instantaneous) may not lead to strictly 
independent components, but then typically it is proposed to use a decomposition algorithm 
within this class which is different from that for truly independent sources \cite{Bach,topica}. 
An exception is the `multidimensional ICA' of \cite{Cardoso98} where the author points out 
that one can use standard decomposition algorithms even in case of non-zero dependencies, 
but also there most of the 
attention is payed to whether components are independent, but not on how dependent they are. 
The latter can be useful for clustering the output, but also for reliability and stability 
testing: A blind source separation into independent components will be the more robust, the 
deeper are the minima of the dependences. In \cite{Meinecke02,noiseinj,icasso} such 
reliability tests have been proposed based on resampling and noise injection. We believe 
that looking at the dependence landscape is more direct and conceptually simpler.

In the present paper we propose to use a recently introduced MI estimator based on 
$k$-nearest neighbor statistics \cite{alex}. It resembles the Vasicek estimator 
\cite{Vasicek} for differential entropies which has been applied recently to ICA 
\cite{Pham,Learned} and which is also based on $k$-nearest neighbor statistics. But while 
the Vasicek estimator exists only for 1-dimensional distributions and can not therefore be 
used to estimate dependencies via MI, our estimator is based on the Kozachenko-Leonenko 
\cite{koza-leon} estimator for differential entropies and works in any dimension. In 
addition, it seems to give the most precise blind source separation algorithm for 
2-d distributions known at present.

Throughout the paper we will only discuss the simplest case of linear superpositions.
While MI can be applied in principle also to nonlinear mixtures, this would be much more 
difficult. 

The paper is organized as follows.
In Sec.~II we recall basic properties of MI and present the MI estimator of \cite{alex}.
The basic version of MILCA is described in Sec.~III where we also give first applications 
to toy models, and where we will also discuss the reliability of the decompositions.
In Sec.~IV we deal with the case where only some groups of output components are independent,
with non-zero interdependencies within the groups. In this case it is natural to cluster
the components. We propose to use again MI for that purpose, in the form of the {\it mutual
information based clustering} (MIC) algorithm presented recently in \cite{alex2}.
In Sec. V we discuss how MILCA (and other ICA algorithms) can be combined with
time delay embedding, in order to take into account non-trivial time structure (in case 
the data to be decomposed form a time series). A thorough discussion of our method and of 
its relations to previous work is given in Sec.~VI.
Conclusions are drawn in the last section, Sec.~VII.

\section{Mutual Information}

\subsection{General Properties of MI}

Assume that $X$ and $Y$ are continuous random variables with joint density $\mu(x,y)$
and marginal densities $\mu_x(x)= \int dy \mu(x,y)$ and $\mu_y(y)$. Then MI is 
defined as \cite{cover-thomas}
\be
   I(X,Y) = \int\!\!\int dx dy \;\mu(x,y) \;\log{\mu(x,y)\over \mu_x(x)\mu_y(y)}\;.
   \label{mi}
\ee

The base of the logarithm determines the units in which information 
is measured. In the following, we will always use natural logarithms, 
i.e. mutual information will be measured in nats.

In terms of the {\it differential entropies} 
\be
   H(X) = - \int dx \mu_x(x) \;\log \mu_x(x), 
\ee
\be
   H(Y) = - \int dy \mu_y(y) \;\log \mu_y(y),
\ee
and
\be 
   H(X,Y) = - \int\!\!\!\int dx dy \;\mu(x,y) \;\log \mu(x,y)
\ee
it can be written as $I(X,Y) = H(X)+H(Y)-H(X,Y)$.

The most important property of MI is that it is always non-negative, and is zero
if and only if $X$ and $Y$ are independent. Another important feature of MI is its 
invariance under homeomorphisms of $X$ and $Y$. 
If $X'=F(X)$ and $Y'=G(Y)$ are smooth and uniquely invertible maps, then
\be 
   I(X',Y') = I(X,Y).
\ee
Notice that this is not the case for differential entropies.
Just as Gaussian distributions maximize the differential entropy, giving thereby 
an upper bound on the entropy in terms of the variance of the distribution, 
Gaussians minimize MI \cite{alex}. This gives a {\it lower} bound on MI
in terms of the correlation coefficient 
\be
   r = {\langle X.Y\rangle \over [\langle X^2\rangle\langle Y^2\rangle]^{1/2}},
\ee
\be
   I(X,Y) \geq -{1\over 2} \log(1-r^2)\;.                     \label{MI-ineq}
\ee

This might suggest that MI can be decomposed into a `linear' part [the r.h.s.
of Eq.(\ref{MI-ineq})] plus a non-linear part. While such a decomposition is 
of course always possible, it is in general not useful. For example, it would 
also suggest that the minimum of MI under linear transformations $(X',Y') = 
{\bf A}\; (X,Y)$ is always reached when $X'$ and $Y'$ are linearly uncorrelated 
(in which case $r=0$ and the r.h.s. of Eq.(\ref{MI-ineq}) is zero). But it is 
easy to give counterexamples for which this is not true (see appendix).

This is important for MILCA, since it is standard practice in ICA to make first a 
``prewhitening" (principle component transformation plus rescaling, so that 
the covariance matrix is isotropic), and to restrict the actual minimization 
of the contrast function to pure rotations \cite{hyvar2001}. If one is sure
that the sources are really independent, this is justified: For the correct 
sources both MIs and covariances are zero. But it is not justified, if there are 
no strictly independent sources and we want to find the {\it least} dependent 
sources.

For any number $M$ of random variables, the MI (or `redundancy', as it is often
called) is defined as
\be
   I(X_1,X_2,\ldots X_M) = \sum_{m=1}^M H(X_m) - H(X_1,X_2,\ldots X_M). 
                           \label{MI-N}
\ee
Notice that this is the appropriate definition for ICA or MILCA, since it 
is this difference which one wants to minimize. In the literature outside 
the ICA community usually a different construct is called MI \cite{cover-thomas},
but we shall in the following only use Eq.(\ref{MI-N}).

The $M$-dimensional MI shares with $I(X,Y)$ the invariance under homeomorphisms
for each $X_m$, and the fact that it is bounded by the value obtained for 
a Gaussian with the same covariance matrix \cite{alex}. 
The next important property is the {\it grouping property} \cite{alex}
\be
   I(X,Y,Z) = I((X,Y),Z) + I(X,Y)\;.        \label{grouping}
\ee
Here, $I((X,Y),Z)$ is the MI between the two variables $Z$ and $(X,Y)$, and we 
have used the fact that a random variable need not be a scalar. Indeed, anything 
we said so far holds also if $X, Y, \ldots $ are multicomponent random variables
(except Eq.(\ref{MI-ineq}) which has to be suitably modified). Therefore, if 
we have more than 3 random variables, Eq.(\ref{grouping}) can be iterated. For 
any set of random variables and any hierarchical clustering of this set into 
disjoint groups, the total MI can be hierarchically decomposed into MIs between 
groups and MIs within each group.
This will become important in Sec.V where we discuss clustering based on MI.

Intuitively, one might expect that $I(X,Y,Z)=0$ if $I(X,Y)=I(X,Z)=I(Y,Z)=0$.
Pairwise strict independence would then imply global independence. That this is 
{\it not} true is demonstrated in the appendix with a simple counter example.
It becomes important for chaotic deterministic systems. If $x_1,x_2,\ldots x_N$
is a univariate signal produced by a strange attractor with dimension $d$, then
any $d$-tuple of consecutive $x_t$ values will be weakly dependent, while any 
$m$-tuple with $m>d$ will be strongly dependent.

The last property to be discussed here is related to homeomorphisms involving a
{\it pair} of variables $(X,Y)$, i.e. $(X',Y') = F(X,Y)$.  Using the
grouping property and the invariance under homeomorphisms of a single variable
we obtain \cite{alex}
\be
   I(X',Y',Z,\ldots) = I(X,Y,Z,\ldots) + [I(X',Y')-I(X,Y)]\;.  \label{dI}
\ee
This is important if we want to minimize the MI with respect to linear
transformations. Since any such transformation in $M$ dimensions can be
factorized into pairwise transformations, this means that we only have to
compute pairwise MIs for the minimization. To find the actual value of the 
minimum, we have of course to perform one calculation in all $M$ dimensions.
We also have to estimate higher order MIs directly, if we want to use the method 
of Sec.~V.A with embedding dimension $m>2$.

\subsection{MI Estimation}

Assume that one has a set of $N$ bivariate measurements, $(x_i,y_i), \, i=1,\ldots N$,
which are assumed to be iid (independent identically distributed) realizations
of the random variable $Z = (X,Y)$. Our task is to estimate MI, with or without
explicit estimation of the unknown densities $\mu(x,y), \mu_x(x)$, and $\mu_y(y)$.

Two classes of estimators were given in \cite{alex}.
In contrast to other estimators based on cumulant expansions, entropy maximalization, 
parameterizations of the densities, kernel density estimators or binnings (for a 
review of these methods see \cite{alex}), the algorithms proposed in \cite{alex} are 
based on entropy estimates from $k$-nearest neighbor distances. This implies
that they are data efficient (with $k=1$ we resolve structures down to
the smallest possible scales), adaptive (the resolution is higher where data are
more numerous), and have minimal bias.
Numerically, they seem to become {\it exact} for independent
distributions, i.e. the estimators are completely unbiased (and therefore
vanish except for statistical fluctuations) if $\mu(x,y) = \mu(x) \mu(y)$.
This was found for all tested distributions and for all dimensions of $x$ and $y$.
It is of course particularly useful for an application where we just want to test
for independence.

In the following we shall discuss only one of these two classes, the one based 
on rectangular neighborhoods called $\hat{I}^{(2)}(X,Y)$ in \cite{alex}.

 \subsection{Formal Developments}

We will start from the Kozachenko-Leonenko estimate for Shannon entropy 
\cite{alex,grass85,somorjai86,koza-leon,victor}:
\be
   \hat{H}(X) = - \psi(k) + \psi(N) + \log c_d + {d\over N} \sum_{i=1}^N \log\epsilon(i)
   \label{KL}
\ee
where $\psi(x)$ is the digamma function, $\epsilon(i)$ is twice the distance from $x_i$
to its $k$-th neighbor, $d$ is the dimension of $x$ and $c_d$ is the volume of the 
$d$-dimensional unit ball. 
Mutual information could be obtained by estimating $H(X)$, $H(Y)$ and $H(X,Y)$ separately
and using 
\be
   I(X,Y) = H(X)+H(Y)-H(X,Y)\;.
\ee
But for any fixed $k$, the distance to the $k$-th neighbor in the joint space will be larger than
the distances to the neighbors in the marginal spaces. Since the bias from the non-uniformity
of the density depends of course on these distances, the biases in $\hat{H}(X)$, $\hat{H}(Y)$,
and in $\hat{H}(X,Y)$ would not cancel.

To avoid this, we notice that Eq.(\ref{KL}) holds for {\it any} value of $k$, and that we 
do not have to choose a fixed $k$ when estimating the marginal entropies (this idea was used
first, somewhat less systematically, in \cite{schreiber}). So let us denote 
by $\epsilon_x(i)$ and $\epsilon_y(i)$ the edge lengths of the smallest rectangle 
around point $i$ containing $k$ neighbors, and let $n_x(i)$ and $n_y(i)$ (the number of 
points with $||x_i-x_j|| \leq \epsilon_x(i)/2$ and $||y_i-y_j|| \leq \epsilon_y(i)/2$) 
as the new number of neighbors in the marginal space. The estimate for MI is then:
\be
   \hat{I}(X,Y) = \psi(k) - 1/k - \langle \psi(n_x)+ \psi(n_y) \rangle + \psi(N).
   \label{i2}
\ee
We denote by $\langle \ldots \rangle$ averages both over all $i\in [1,\ldots N]$ and
over all realizations of the random samples.

\begin{figure}
  \begin{center}
    \psfig{file=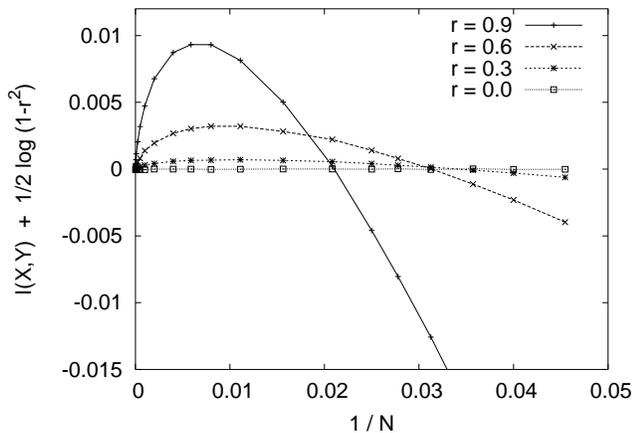,width=6.0cm,angle=270}
 \caption{Estimates of average values of $I(X,Y) - I_{\rm exact}(X,Y)$ for Gaussian
   signals with unit variance and covariances $r=0.9, 0.6, 0.3$, and 0.0 (from top to 
   bottom), plotted against $1/N$. In all cases $k=1$. The number of realizations is
   $>2\times 10^6$ for $N<=1000$, and decreases to $\approx 10^5$ for $N=40,000$.
   Error bars are smaller than the sizes of the symbols.}
 \label{gauss}
 \end{center}
\end{figure}

Here we will show results of $\hat{I}(X,Y)$ for Gaussian distributions (cf. Fig.~\ref{gauss}). 
Let $X$ and $Y$ be Gaussian signals with zero mean and unit variance, and with covariance 
$r$. In this case $I(X,Y)$ is known exactly,

\be
   I_{\rm Gauss}(X,Y) = -{1\over 2} \log(1-r^2)
   \label{gauss-MI}
\ee
Apart from the fact that indeed $\hat{I}(X,Y) - I_{\rm Gauss}(X,Y) \to 0$ for $N\to \infty$, 
the most conspicuous feature is that the systematic error is compatible with zero for $r=0$.
This is a property which makes the estimator particular interesting for ICA because there 
we are looking for uncorrelated signals. For non-Gaussian signals, our estimator still 
has a smaller systematic error than other estimators in the literature \cite{alex}.

Using the same arguments for $n$ random variables $X_1,X_2, \ldots X_m$, the MI estimate 
for $I(X_1,X_2, \ldots X_m)$, is \cite{alex}:
\bea
   \hat{I}(X_1,X_2, \ldots X_m)& = & \psi(k) - (m-1)/k + (m-1)\psi(N)  \nonumber \\
    & - & \langle \psi(n_{x_1}) + \psi(n_{x_2}) + \ldots \psi(n_{x_m}) \rangle \nonumber
\eea

%Fig.1

 \subsection{Practical considerations}

By choosing proper values for $k$, the algorithm allows to minimize either the statistical
or the systematic errors. The higher is $k$, the lower is the statistical error of $\hat{I}$.
The systematic error shows exactly the opposite behavior. Thus, to keep the balance between 
these two errors, the best choice for $k$ would lie in the middle range. But for some cases 
it makes sense to deviate from this, e.g. when we want to find most independent signal sources.
There the true values of the MI are small, and thus also the systematic errors for all $k$.
In this case it is better to use large $k$ in order to reduce statistical errors. On the 
other hand, when the data files are very long we do not have to worry about statistical errors 
and we should choose $k$ small. 

Most of the CPU time for estimating MI with our new estimator is used for neighbor 
searching. In \cite{alex} we presented three implementations which ranged from very simple
but slow to sophisticated and fast. In the following we shall always use the fastest
implementation which uses grids to achieve a CPU time $\sim N\log N$ for $N$ points. We 
will not use rank ordering (as also discussed in \cite{alex}), but we will add small 
Gaussian jiggles (amplitude $\approx 10^{-8}$) to all measured values in order to break 
any degeneracies due to quantization in the analog-to-digital conversion \cite{alex}.

\section{MILCA without using Temporal Structures}

 \subsection{Basic Algorithm}

In this Subsection we will show how the linear instantaneous ICA problem is solved using 
the new MI estimator. We will apply this then to several artificial data sets which are 
constructed by superimposing known independent sources, and we will compare the results with 
those from several other ICA algorithms.

In the simplest case ${\bf x}(t)$ is an instantaneous linear superposition of $n$
independent sources ${\bf s}(t)=(s_1(t),s_2(t),...,s_n(t))$,
\be
  {\bf x}(t) =  {\bf A}\;{\bf s}(t)\; ,
  \label{source-x}
\ee
where ${\bf A}$ is a non-singular $n\times n$ `mixing matrix'. This means that the number of 
sources is equal to the number of measured components.
In this case, we know that a decomposition into independent components is
possible, since the inverse transformation
\be
   {\bf \hat{s}}(t) = {\bf Wx}(t) \;\;\text{ with } \;\; {\bf W}={\bf A}^{-1}
  \label{x-source}
\ee
does exactly this. If Eq.(\ref{source-x}) does not hold then no decomposition
into strictly independent components is possible by a linear transformation
like Eq.(\ref{x-source}), but one can still search for least dependent
components.

But even if Eq.(\ref{source-x}) does hold, the problem of blind source
separation (BSS), i.e., of finding the matrix ${\bf W}$ without explicitly knowing
${\bf A}$, is not trivial. Basically, it requires that ${\bf x}$ is such that
the components of any superposition ${\bf s}' = {\bf W}'{\bf x}$ with ${\bf W}'\ne {\bf W}$
are not independent. Since linear combinations of Gaussian variables are also
Gaussian, BSS is possible only if the sources are not Gaussian. Otherwise, any
rotation (orthogonal transformation) ${\bf s}' = {\bf Rs}$ would again lead
to independent components, and the original sources ${\bf s}$ could not be
uniquely recovered. Since any ICA algorithm will find a more or less meaningful
solution, we need a reliability test for the obtained components. This is given 
in subsection C.

As a first step, the matrix {\bf W} is usually decomposed into two factors, 
${\bf W} = {\bf R}{\bf V}$, where the {\it prewhitening} {\bf V} transforms the 
covariance matrix into ${\bf C}' = {\bf V}{\bf C}{\bf V}^T = {\bf 1}$, and {\bf R} 
is a pure rotation. Prewhitening is just a principal component analysis (PCA) 
together with a rescaling. The ICA problem proper reduces then to finding
a suitable rotation for the prewhitened data. 

The motivation for this is that 
any reasonable contrast function used for the ICA will give least dependent components
which are also uncorrelated. In Sec.~II we have seen that this is not always the case,
but that it is true whenever the components are really {\it independent}.
One can take now several different attitudes. The most radical is to abandon
prewhitening altogether (for different reasons for not to use prewhitening, see 
\cite{ziehe03}). But this slows down the algorithm considerably. Also, prewhitening
can be detrimental only when there are residual dependencies between the optimal
components, and it is not clear what is the significance of such components. In the 
following we shall always use prewhitening unless we say explicitly the opposite.
We shall always assume that the prewhitening step has already been done, and we will
restrict the proper ICA (or rather LCA) transformations to pure rotations. As a third
alternative one could first use prewhitening, but try at the end whether some 
non-orthogonal transformations improve the results further. We have not yet studied 
this strategy.

\begin{table*}
\begin{center}
\label{table:comp}
\begin{ruledtabular}
\begin{tabular*}{7.cm}{||c|c|c|c|c|c|c|c|c||}
\hline
pdfs & FastICA & Jade &  Imax  & KCCA &  KGV  & RADICAL & MILCA & MILCA(augmented) \\ \hline
  a  &   4.4   &  3.7 &\bf{1.8}&  3.7 &  3.0  &   2.1  &   2.7  &   2.4    \\
  b  &   5.8   &  4.1 &   3.4  &  3.7 &  2.9  &   2.7  &   2.9  & \bf{2.5} \\
  c  &   2.3   &  1.9 &   2.0  &  2.7 &  2.4  &   1.2  &   1.5  & \bf{1.0} \\ \hline
  d  &   6.4   &  6.1 &   6.9  &  7.1 &  5.7  &   5.3  &   7.0  & \bf{4.3} \\
  e  &   4.9   &  3.9 &   3.2  &  1.7 &  1.5  &\bf{0.9}&\bf{0.9}&   1.0    \\
  f  &   3.6   &  2.7 &   1.0  &  1.7 &  1.5  &   1.0  &\bf{0.9}& \bf{0.9} \\ \hline
  g  &   1.8   &  1.4 &\bf{0.6}&  1.5 &  1.4  &\bf{0.6}&\bf{0.6}& \bf{0.6} \\
  h  &   5.1   &  4.1 &\bf{3.1}&  4.6 &  3.6  &   3.7  &   3.4  &   3.3    \\
  i  &  10.0   &  6.8 &   7.8  &  8.3 &\bf{6.4}&  8.3  &   7.9  &   8.0    \\ \hline
  j  &   6.0   &  4.5 &  50.6  &  1.4 &  1.3  &   0.8  &\bf{0.7}&   0.8    \\
  k  &   5.8   &  4.4 &   4.2  &  3.2 &  2.8  &   2.7  &   2.4  & \bf{2.3} \\
  l  &  11.0   &  8.3 &   9.4  &  4.9 &  3.8  &   4.2  &   4.1  & \bf{3.3} \\ \hline
  m  &   3.9   &  2.8 &   3.9  &  6.2 &  4.7  &   1.0  &   1.0  & \bf{0.8} \\
  n  &   5.3   &  3.9 &  32.1  &  7.1 &  3.0  &   1.8  &   2.0  & \bf{1.6} \\
  o  &   4.4   &  3.3 &   4.1  &  6.3 &  4.5  &   3.4  &   3.4  & \bf{2.9} \\ \hline
  p  &   3.7   &  2.9 &   8.2  &  3.6 &  2.8  &\bf{1.1}&   1.6  &   1.2    \\
  q  &  19.0   & 15.3 &  43.3  &  5.2 &  3.6  &   2.3  &   2.9  & \bf{1.9} \\
  r  &   5.8   &  4.3 &   5.9  &  4.1 &  3.7  &   3.2  &   3.5  & \bf{2.7} \\ \hline
mean &   6.1   &  4.7 &  10.6  &  4.3 &  3.3  &   2.6  &   2.7  & \bf{2.3} \\ \hline
\end{tabular*}
\end{ruledtabular}
\caption{Performance indices (multiplied by 100) for two-component blind source separation, 
test problem (A).
The results in the first 6 columns (FastICA, Jade, Imax, KCCA, KGV, and RADICAL) are
taken from Ref.\cite{Learned}, where also references to these algorithms are given and
where the probability distribution functions (pdfs) `a' - `r' are defined. The last two
columns show the results of MILCA, first in its simplest version (column 7) and then with
data augmentation as proposed in \cite{Learned} (column 8). Each performance index is an
average over 100 replicas, each replica consisting of 1000 pairs of numbers drawn randomly
from the pdfs. For MILCA we used $k=10$, and we fitted $\hat{I}(\phi)$ by Fourier sums with 3
(MILCA) and 5 terms (augmented MILCA), respectively.}
\end{center}
\end{table*}

%Figure 2
\begin{figure}
  \begin{center}
    \psfig{file=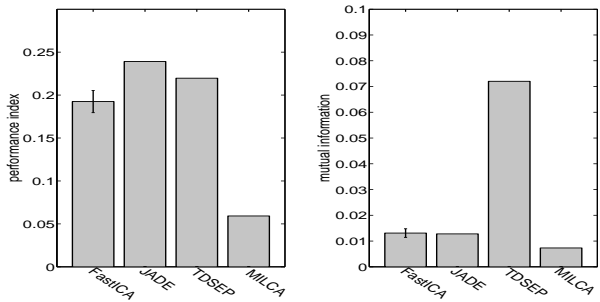,height=4cm,width=8cm,angle=0}
%   \vskip 0.3 true cm
 \caption{Test problem (B), consisting of 5 input channels.
    Left panel: Averaged performance index $P_{err}$ from the output of
    FastICA \cite{hyvar2001} (parameters with lowest MI), JADE \cite{Cardoso93},
    TDSEP (same parameters as in \cite{Meinecke02}), and MILCA (k=30).
    Right panel: same as left side, but with total MI $\hat{I}$ (k=3) used as performance measure.}
 \label{algocomp}
 \end{center}
 \end{figure}

%Figure 3
\begin{figure}
  \begin{center}
    \psfig{file=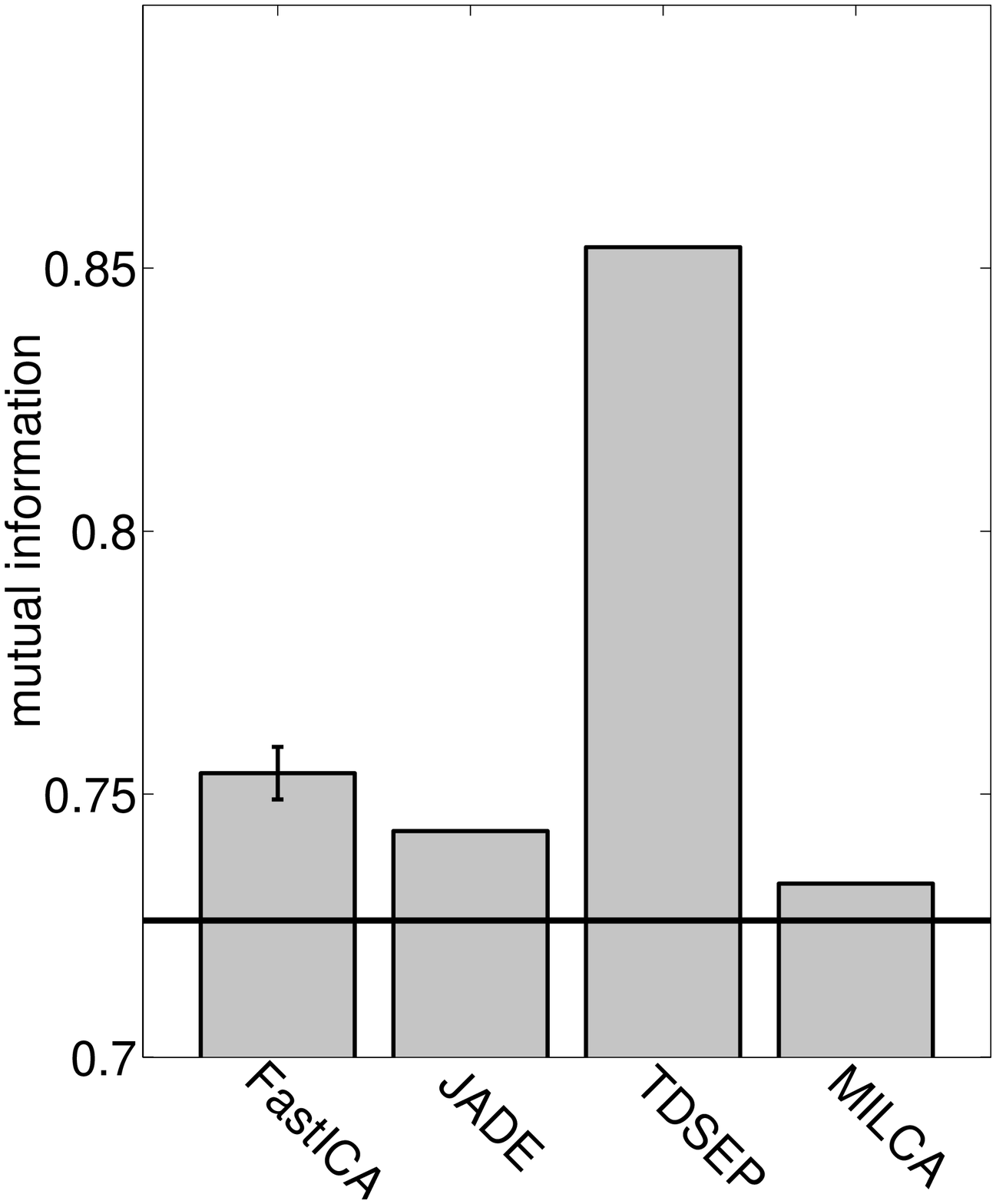,width=3.95cm, height=3.8cm, angle=0}
    \psfig{file=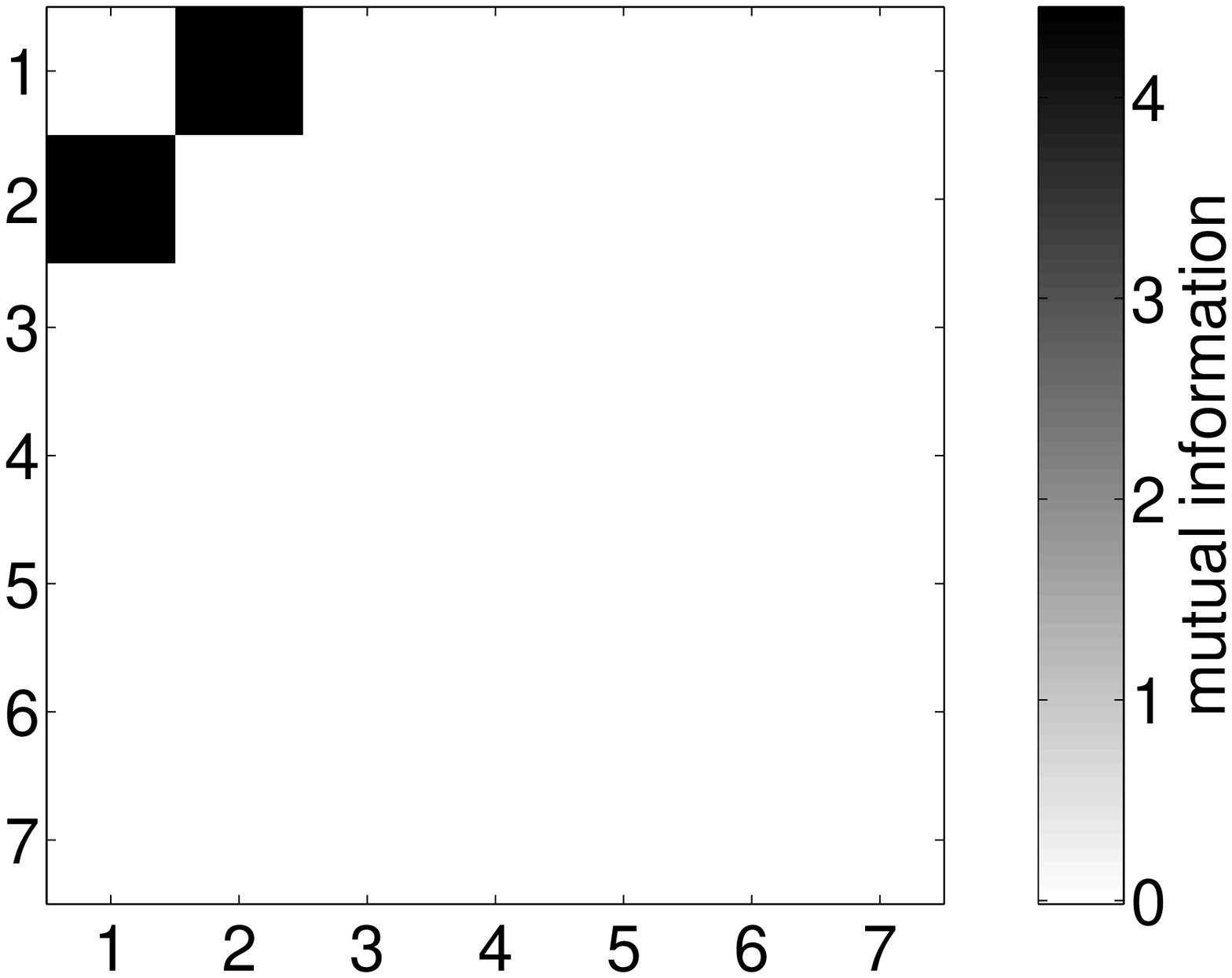,width=3.95cm, height=3.8cm, angle=0}
 \caption{Test problem (C), with 7 input channels.
   Left panel: Averaged $\hat{I}(\hat{s}_1\ldots \hat{s}_n)$ (k=3) from the output of 
   FastICA \cite{hyvar2001} (parameters with lowest MI), JADE \cite{Cardoso93},
   TDSEP (same parameters as in \cite{Meinecke02}), and MILCA (k=30). The horizontal
   line indicates the true MI of the input channels.
   Right panel: Pairwise MI estimates $\hat{I}$ between all channel combinations,
   for the MILCA output components shown in Fig. \ref{MILCAoutput}
   (diagonal is set to zero).}
 \label{algocompdep}
 \end{center}
 \end{figure}

The aim of ICA is now to minimize $I(X_1\ldots X_n)$ under a pure rotation {\bf R}.
Any rotation can be represented as a product of rotations which act only in some
$2\times 2$ subspace, ${\bf R} = \prod_{i,j} {\bf R}_{ij}(\phi)$, where
\be
   {\bf R}_{ij}(\phi)(x_1\ldots x_i\ldots x_j\ldots x_n) = (x_1\ldots x_i'\ldots x_j'\ldots x_n)
   \label{R}
\ee
with
\be
   x_i' = \cos\phi\; x_i + \sin\phi\; x_j,\quad x_j' = -\sin\phi\; x_i + \cos\phi\; x_j\;.
\ee
For such a rotation one has (see Eq.(\ref{dI}))
\be
   I({\bf R}_{ij}(\phi){\bf X}) - I({\bf X}) = I(X_i',X_j') - I(X_i,X_j)\;,
   \label{change-I}
\ee
i.e., the change of $I(X_1\ldots X_n)$ under any rotation can be computed by adding up
changes of two-variable MIs. This is an important numerical simplification.

To find the optimal angle $\phi$ in a given $(i,j)$ plane, we calculated $\hat{I}_{ij}(\phi)
 = \hat{I}(X_i',X_j')$
for typically 150 different angles in the interval $[0,\pi/2]$, fitted these values
by typically 3-15 Fourier components, and took then the minimum of the fit. The latter 
is useful because $\hat{I}(\phi)$ is not smooth in $\phi$, for essentially the same reasons 
as discussed in \cite{Learned}. We also tried the augmentation proposed in \cite{Learned}
to smoothen $\hat{I}(X',Y')$. It worked equally well, by and large, as the Fourier filtering,
but it was much slower.

Now the resulting MILCA-algorithm can be summarized:

\begin{enumerate}
  \item Preprocess (center, filter, detrend, ...) and whiten the data;
  \item For each pair $(i,j)$ with $i,j = 1\ldots n$ find the angle $\phi$ which minimizes a 
        smooth fit to $\hat{I}_{ij}(\phi) = \hat{I}(X_i',X_j')$;
  \item If $\hat{I}(X_1'\ldots X_n')$ has not yet converged, go back to step 2. Else, 
        $\hat{s}_i = X_i'$ are the estimates for the sources.
\end{enumerate}

The order of choosing the sequence of pairs in point 2 is not essential. In our numerical 
simulations the convergence speed did not differ significantly whether we went through the 
pairs $(i,j)$ systematically or randomly.

\subsection{Numerical Examples and Performance Tests}

%Figure 4
\begin{figure}
  \begin{center}
    \psfig{file=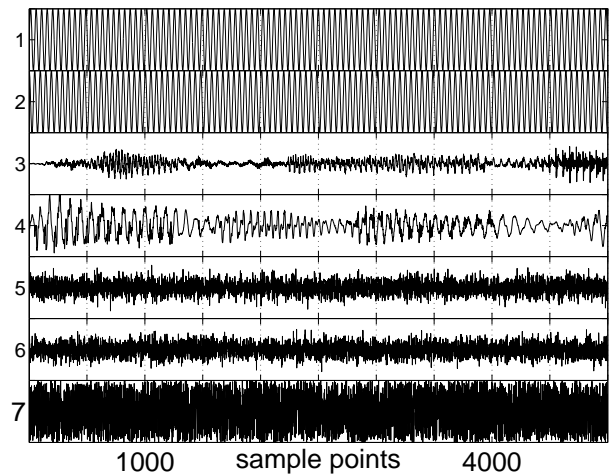,width=8.0cm,angle=0}
%   \vskip 0.3 true cm
 \caption{Seven output channels of the MILCA algorithm, test problem (C).}
 \label{MILCAoutput}
 \end{center}
\end{figure}

{\bf (A)} As a first test we study the set of 18 problems proposed by Bach \& Jordan \cite{Bach2002}
and studied also in \cite{Learned}. Each problem corresponds to a 1-d probability distribution
$p(x)$. One thousand pairs of random numbers $x$ and $y$, each drawn iid from $p(x)p(y)$, are 
mixed as $x'=x \cos\phi + y \sin\phi, y' = - x \sin\phi + y \cos\phi$ with random angle $\phi$
common to all pairs (i.e., ${\bf A}$ is a pure rotation). Using MILCA, we obtained then the 
estimate $\hat{\bf A}$. This is repeated 100 times with different angles $\phi$ and with 
different random sets of pairs $(x,y)$. To assess the quality of the estimator $\hat{\bf A}$
(or, equivalently, of the back transformation $\hat{\bf W} = \hat{\bf A}^{-1}$), we use 
the Amari performance index $P_{err}$ \cite{Amari}
\be
  P_{err} = {1\over 2N} \sum_{i,j=1}^N ({|p_{ij}| \over \max_k|p_{ik}|} +
            { |p_{ij}| \over \max_k|p_{kj}|})-1
  \label{perfind}
\ee
where $p_{ij} = (\hat{\bf A}^{-1}{\bf A})_{ij}$.

Results are given in Table~1 (column `MILCA') and compared there to the results of previous 
algorithms given in \cite{Learned}. They are excellent in average and surpassed only by 
the RADICAL algorithm proposed in \cite{Learned} which also uses an entropy estimate based
on neighbor distances, but for the differential Shannon entropies $H(x')$ and $H(y')$.
Another feature used in \cite{Learned} is {\it data augmentation}: To obtain a more smooth
dependence on the angle $\phi$, each data vector $(x,y)$ is replaced by an R-tuple (with 
$R=30$) of near-by points. The same augmentation trick can be used also for MILCA, and 
improves the results for very similar reasons. Indeed, our results obtained with MILCA and 
with data augmentation, given in the last column of Table~1, are even better than those 
of RADICAL. In the following tests we did not use data augmentation, because it is rather 
time consuming.

{\bf (B)} As a second test we study an example taken from \cite{Meinecke02}. In involves five input 
sources (a sine wave, two different speech signals [the first half of ``Houston, we have a problem" 
and ``parental guidance is suggested" from \cite{audiosignal}], one white Gaussian noise, one 
uniformly distributed white noise) (5000 data points each) which are linearly mixed with a
$5\times 5$ matrix {\bf A} to form five output signals.
In mixing these components, no time delay is used, i.e. the superpositions are strictly local
in time. For this example it is possible to find the inverse transformation ${\bf W} = 
{\bf A}^{-1}$ up to a permutation and up to scaling factors, because all sources are 
independent from each other and only one has a Gaussian distribution.
To assess the quality of this back transformation we again use the Amari performance index.

The results obtained with two hundred different random mixtures of the sources (with uniformly 
distributed mixing matrices and with different realizations of the random channels for 
each mixture) are compared  in the left panel of Fig.~\ref{algocomp} with three standard 
algorithms: FastICA~\cite{hyvar2001}, JADE~\cite{Cardoso93}, and TDSEP~\cite{Meinecke02}.
We found that FastICA sometimes gets stuck in a local minimum, and runs differing only in the 
initial conditions can produce different results. The error bars shown in Fig.~\ref{algocomp} 
indicate the resulting uncertainty of the performance measure, estimated from 20 realizations 
that differ only in initial conditions.
The errors of JADE and FastICA are mainly due to their difficulty to separate one of the audio 
channels from the Gaussian noise. TDSEP is not able to decompose the two noise channels, since
it is also not designed for this purpose (it uses time structures to separate signals). Very 
good results for all 200 mixtures are obtained by MILCA, although the audio signals are 
quit noisy and have nearly Gaussian distributions. The performance of JADE and FastICA 
compared to MILCA becomes better when the quality of the acoustic signals improves.

In addition to the Amari index, another (more direct) way to judge the accuracy of the source
estimates is to look at the estimated MIs. If and only if the sources were estimated correctly, the 
MI should be zero. In the following we propose to use both the matrix of pairwise estimators 
$\hat{I}(\hat{s}_i,\hat{s}_j)$ and the estimated total MI $\hat{I}(\hat{s}_1\ldots \hat{s}_n)$.  
The important advantage over the Amari index is that they can also be used when the exact sources 
are not known. Low values of the MI indicate that {\it both} the data is a mixture of independent
components, {\it and} the separation algorithm worked well in producing {\it some} independent
components. Notice that it cannot be expected in general that the components found are identical
to the sources, e.g., if some of them are Gaussians. In Fig.~\ref{algocomp} again MILCA 
shows the best performances.

Notice the very big difference between FastICA/JADE and TDSEP in the right panel of 
Fig.~\ref{algocomp}, which is much bigger than that measured with the Amari index.
The first two have problems in separating one of the acoustic signals (signal \#4 in 
Fig.~\ref{MILCAoutput}) from the Gaussian, because it has a nearly Gaussian amplitude 
distribution, but for the same reason this is not punished 
by a large MI between the outputs (improved performance index see later in Fig.~\ref{perfdelay}).
TDSEP, using time information, has no problem with this, but cannot separate uniform
from Gaussian noise -- and is heavily punished for that by MI. In
Sec.~V we will show how to improve MILCA such that it can better separate components which
have nearly Gaussian amplitude distributions but different time correlations. Using that 
improved MILCA will give much bigger performance difference with algorithms like FastICA/JADE.

{\bf (C)} Next we want to investigate the case where the decomposition is neither perfectly nor
uniquely possible. Such an example can be constructed by simply adding one cosine with the same 
frequency as the sine and one more Gaussian channel to the last test case.
This now violates the assumption of independent sources, because the sine and cosine are strongly
dependent. The theoretical value for the MI would be infinite, but a numerical estimator from a
finite data sample gives a finite value, in our case  $I(S_1\ldots S_n)=0.72$ \cite{footnote3}.
But for this example,
perfect blind source separation is impossible also because the two Gaussians are not uniquely 
decomposable. We want to know how an ICA algorithm performs in view of such problems. It should
still be able to separate those components which can be separated.

The total output MI is shown in the left panel of Fig.~\ref{algocompdep}. We see that for all 
algorithms the MI is higher than the MI between the input channels, which serves essentially as 
a consistency test. The difference is smallest for MILCA. The MIs between all pairwise channel 
combinations obtained with MILCA are shown in the right panel of Fig.~\ref{algocompdep}. They 
show again that MILCA has done a perfect job: All components are independent except for those 
which should not be. MILCA output is shown directly in Fig.~\ref{MILCAoutput}. Although we do 
not show the input, it is clear that the separation has been as successful as possible.

{\bf (D)} There are a number of blind source separation problems in the field of
analytical spectroscopy, where quantitative spectral analysis of chemical mixtures 
is formulated as {\it multivariate curve resolution} (for recent reviews see
\cite{smcr,geladi2003,tauler2003} and as an ICA problem \cite{chemica, epr, chemica2004}).
Assuming Beer's law, the spectrum of a mixture of pure constituents with spectra 
$s_i(\nu)$ and concentrations $A_i$ is $x(\nu) = \sum_i A_is_i(\nu)$. Given a set 
of $N$ mixtures and $N$ pure components, we can then write this in vector notation 
as ${\bf x}(\nu)={\bf A}{\bf s}(\nu)$, analogous to Eq.(\ref{source-x}).
The task is to obtain estimates ${\bf \hat s}(\nu)$ for the pure components.
This is the instantaneous linear ICA problem, except that in most applications
of interest the spectral sources are not independent but have overlapping bands.
This happens when chemical compounds in a mixture share several common or similar 
structural groups that demonstrate nearly the same spectral patterns.

%Figure 5
\begin{figure}
  \begin{center}
    \psfig{file=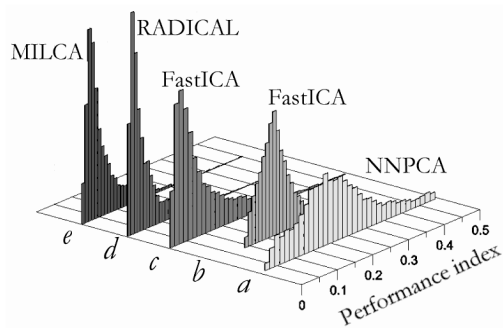}
    \caption{Performance index distributions over 7000 triples of three-component 
     mixtures. For histograms {\it (a),(b)} the original spectra were decomposed, 
     for {\it (c)-(e)} their second derivatives.}
 \label{spectral}
 \end{center}
 \end{figure}

This difficulty makes mixture decomposition quite nontrivial for many BSS
techniques used in chemometrics, unless interactive band selection 
(e.g. SIMPLISMA \cite{simplisma}, IPCA \cite{ipca}, BTEM \cite{btem}) is employed
to avoid using those parts of the signals where severe overlaps reduce the
quality of decomposition. Such preprocessing made by hand is, of course, a bit
of an art, because these unsafe bands can not be known {\it a priori} in a
blind problem. Since the focus here is rather on developing general purpose
algorithms, we aim at using MILCA without interactive preprocessing in order
to estimate its pure overall efficiency in cases when residual dependencies
play a role.

To test the performance of MILCA on typical spectral data we collected a pool
of 62 experimental molecular infrared absorption spectra in the range
550-3830 $cm^{-1}$ (822 data points each) taken from the NIST database
\cite{nist}. This test set
was selected to contain organic compounds with common structural groups
(benzene derivatives, phenols, alcohols, thiols) so that their spectra have
multiple overlapping bands and, thereby, are mutually dependent. Then a 
sample of 7000 triples of three-component mixtures was constructed by choosing 
spectra randomly from the pool and applying random mixing matrices {\bf A}
\cite{footnote5}.
For each decomposition the Amari performance index was computed. Fig.~\ref{spectral}
compares its distributions for several different ICA algorithms including 
FastICA \cite{hyvar2001}, RADICAL \cite{Learned} and Nonnegative PCA (NNPCA)
\cite{nnpca}. The latter uses the fact that pure spectra are non-negative and
the same should hold for the estimates, so the nonnegativity is imposed as a 
soft constraint on the estimates ${\hat s_{\it i}}(\nu)$ in an optimization 
procedure. But our simulations showed that this constraint is often not 
fulfilled, and in some cases the output of NNPCA (as well as that of other 
algorithms) is negative. To a large part this is due to dependencies between
the sources. Already prewhitening (i.e. PCA and rescaling) sometimes leads to 
decorrelated components which cannot be made nonnegative by any subsequent 
rotation. Trying to enforce nonnegativity neglecting other aspects
might then be counterproductive, and this might partly explain
the relatively poor performance of NNPCA (Fig.~\ref{spectral}{\it a}).

NNPCA has to be applied to the original spectra, while it is well known
that using derivatives of spectroscopic signals with respect to frequency can 
improve the results (see, e.g., \cite{chemica,chemica2004}). Taking such
derivatives extracts the spectral information
which is more independent between the sources \cite{Amaribook}. In our numerical
experiments, second order derivatives approximated by finite differences
\be
  \left.{d^{2}{\bf x}(\nu) \over d{\nu}^2} \right|_{{\nu}_i} \sim {\bf x}({\nu}_{i-1}) -
  2 {\bf x}({\nu}_{i}) + {\bf x}({\nu}_{i+1}).
  \label{diff}
\ee
gave best performance \cite{footnote6}. This is clearly 
seen in the example of FastICA (compare distributions {\it (b)} and {\it(c)} 
on Fig.~\ref{spectral}). But MILCA {\it (e)} and RADICAL {\it (d)} with second 
derivative data perform better than FastICA {\it (c)}, and are almost equally 
good when compared to each other. Furthermore, our numerical results confirmed
that nonnegativity is satisfied whenever the decomposition is successful 
(Amari index below 0.05) (see also the discussion in \cite{cich}). But whether 
this is fulfilled depends primarily on the dependencies between the original 
signals, and less on the algorithm employed.

A more detailed study of the potential of MILCA in multivariate spectral curve
resolution will be given in a forthcoming publication \cite{astakh} which will 
focus on the analysis of experimental mixtures and, in particular, on the comparison 
with recently developed interactive algorithms such as BTEM \cite{btem}.

\subsection{Reliability and Uniqueness of the ICA Output}

%Figure 6
\begin{figure}
  \begin{center}
    \psfig{file=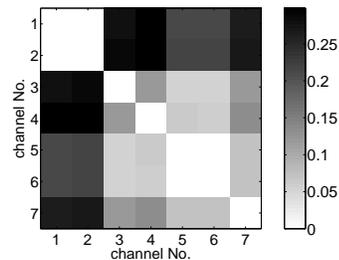,width=4.4cm,angle=0}
 \caption{Square roots of variabilities ${\sigma_{ij}}$ of
      $I(R(X_i, X_j))$ (with $k=6$) from MILCA output for test problem (C)
      (Fig.~\ref{MILCAoutput}). Elements on the diagonal have been set to zero.}
 \label{figvar}
 \end{center}
 \end{figure}

Obtaining the most independent components from a mixture is only the first part of an ICA analysis.
Checking the actual dependencies between the obtained components should be the next task, although 
it is most often ignored. We have seen that it becomes easy and natural with MILCA, which was 
indeed one of our main motivations for MILCA.
The next task after that is to check the reliability, uniqueness, and robustness of the decomposition. 
We have already discussed this in the last subsection for test example (C), but not very systematic.
A systematic discussion will be given now.

Recently proposed reliability tests \cite{Meinecke02,noiseinj,icasso} are based on 
bootstrap methods or noise injection. We here present an alternative procedure which again makes
use of the fact that MILCA gives reliable estimates of the {\it actual (in-)dependencies}:
We test how much the estimated dependencies change under re-mixing the outputs.

In the simplest case, a multivariate signal with $n$ components is an instantaneous linear mixture 
of $n$ independent sources. This was the model we started with in subsection A. We assume it to 
apply when (i) all
estimated pairwise MIs between all ICA components fall below a defined threshold,
$\hat{I}(\hat{s}_i,\hat{s}_j)<D_{\rm max}$ for all $i,j=1,...,n$  and $i\neq j$, and (ii) the overall
MI $\hat{I}(\hat{s}_1\ldots\hat{s}_n)$ is below another threshold. Notice that the first criterion
alone is not sufficient, see the appendix.

In real-world data, however, we are usually confronted with deviations from this simple model.
The next simple possibility is that some pairwise MIs are still exactly zero, but others are not.
Let us draw a graph where each of the $n$ output channels is represented by a vertex, and 
each pair $(i,j)$ of vertices is connected by an edge if $\hat{I}(\hat{s}_i,\hat{s}_j) > D_{\rm max}$.
This gives a partitioning of the set of output components into connected clusters $C_1,\ldots
C_m$ with $m\leq n$. If, in addition, the MI between these clusters, $\hat{I}(C_1,\ldots C_m$), is 
below another suitably chosen threshold, we consider each cluster to be independent (notice that
we do not require all channels within a cluster to have a MI above the threshold $D_{\rm max}$).
This is essentially our version of multidimensional ICA \cite{Cardoso98}. It uses exactly the 
same basic MILCA algorithm as defined above, and is thus much simpler conceptually than the 
`tree-dependent component analysis' of \cite{Bach}. Its main drawback is that it is not sensitive
to the actual strengths of the non-zero interdependencies. A better algorithm which does take them
into account will be discussed in Sec.IV.

In addition to this first step of an ICA output analysis, we have to test for the uniqueness of the 
components. For this purpose, we check whether the (one- or multi-dimensional) sources obtained
by the ICA algorithm indeed correspond to distinct minima of the contrast function or whether 
other linear combinations exist which show approximately the same overall dependencies.
An example for the latter case is given by two uncorrelated Gaussian signals. They
remain independent under rotation \cite{foot7}.

A good estimator for the uniqueness of the ICA output is the variability of the pairwise 
MI under remixing, i.e. under rotations in the two-dimensional plane:
\begin{equation}
   \sigma_{ij} =   \overline{I(X_i,X_j)} - \hat{I}_{ij}(\phi_{\rm min}) \quad {\rm for} \;\;i\neq j
             \label{eqvar}
\end{equation}
where the global minimum of $\hat{I}$ is at $\phi=\phi_{\rm min}$, and
\begin{equation}
   \overline{I(X_i,X_j)} = {2\over \pi}\int_0^{\pi/2} d\phi \;\hat{I}_{ij}(\phi)
\end{equation}
(notice that $I_{ij}(\phi)$ is periodic in $\phi$ with period $\pi/2$).
For unique solutions the MI will change significantly (large $\sigma_{ij}$), but it will 
stay almost constant for ambiguous outputs (small $\sigma_{ij}$).

Results for the MILCA output of test problem (C) are shown in Fig.~\ref{figvar}. (to aid in the 
interpretation, the actual output signals were shown in Fig.~\ref{MILCAoutput}). The basic ICA 
model is violated both in the Gaussian noise subspace and the $sin/cos$ subspace. In the 
Gaussian subspace the components are independent, but it should be impossible to find a unique 
decomposition. Indeed, $\sigma_{5,6}\approx 0$ (Fig~\ref{figvar}) and $\hat{I}_{5,6}\approx0$
(Fig.~\ref{algocompdep}). For the dependent components ($sin/cos$ subspace) the situation is different. 
We expect to have $\sigma= 0$ also here, corresponding to the isotropy of the distribution in this 
subspace. But $\hat{I}$ should be much larger than zero, because the two signals are not independent.
Indeed, we see $\sigma_{1,2}\approx 0$ and $\hat{I}_{1,2}\gg0$. In general, it depends on the 
specific application whether one should attribute any meaning to $\sigma_{ij}$ when components 
$i$ and $j$ are not independent. Finally, we conclude from Fig.~\ref{algocompdep}(right) and 
Fig.~\ref{figvar} that the channels 3, 4 (audio signals), and 7 (uniformly distributed noise) are 
one dimensional sources, because they are independent from any channel, $\hat{I}_{3,i}\approx 
\hat{I}_{4,i}\approx \hat{I}_{7,i}\approx0$, and are reliable, $\sigma_{3,i}\approx\sigma_{4,i}
\approx\sigma_{7,i}\gg0$.

%Figure 7
\begin{figure}
  \begin{center}
    \psfig{file=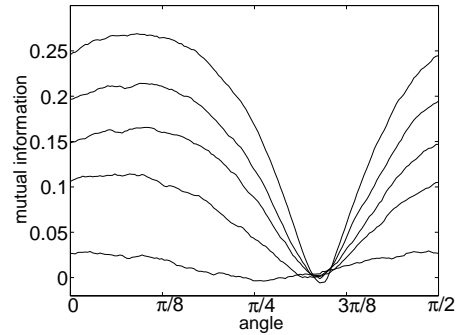,width=6.0cm,angle=0}
 \caption{Unsmoothened estimates of $\hat{I}(\phi)$ for two randomly mixed uniform
   distributions, corrupted with isotropic Gaussian measurement noises with different
   signal-to-noise ratios $SNR = \infty,13,7,4,1$ (from top to bottom), plotted against $\phi$.}
 \label{noiseangle}
 \end{center}
 \end{figure}

%Figure 8
 \begin{figure}
  \begin{center}
    \psfig{file=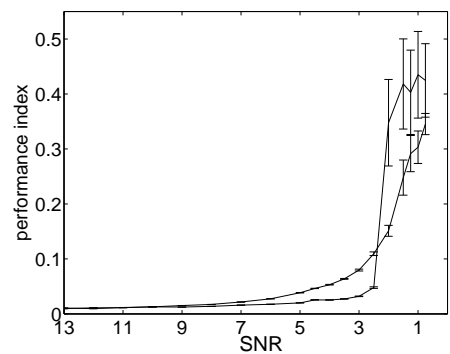,width=6.0cm,angle=0}
 \caption{Averaged Amari index against the signal-to-noise ratio. The condition number of the
   mixing matrices is 6. The upper curve (in the SNR range from 7 to 3) is for standard MILCA, the lower for n-MILCA.}
 \label{audstn}
 \end{center}
 \end{figure}

 \subsection{Noisy Signals}

Because our aim is to apply MILCA to real world data, we have to discuss the influence of
measurement noise. In the literature there exist several algorithms which are specially tailored
to this problem (see e.g. Ref.~\cite{hyvar2001}, chapter 15). Typically, in order to obtain
optimal performance, the noise is assumed to satisfy very special properties like being additive,
uncorrelated, isotropic and Gaussian. Below we will present a modified MILCA algorithm which
assumes that we have measurement noise with exactly these properties.

Alternatively, one can take just a standard ICA algorithm (in our case MILCA as described 
above), and analyze how its output depends on the noise level. In the following we will 
compare both approaches.

We start with two uniformly distributed variables and mixed them with a random $2\times 2$
matrix with a fixed condition number. After that, iid Gaussian noises
are added to each of the two mixtures. The amplitudes in both channels are the same,
\be
   x_i(t) = \sum_{j=1}^2 A_{ij} s_j(t) + \eta_i(t)
\ee
with $\langle \eta_i(t) \eta_j(t')\rangle = r \delta_{ij} \delta_{tt'}$. For the case were
we do not use any information of the measurement noise signals $x_i(t)$
are then simply used as input in MILCA. In Fig.~\ref{noiseangle} we show $\hat{I}(\phi)$
for the same mixing matrix but different signal-to-noise ratios $SNR={\rm var}(s_i(t))/r$.
We see that $\hat{I}$ becomes flatter (the variability with respect to
the mixing angle decreases) with decreasing SNR \cite{footnote}. The presence
of noise leads also to a shift of the minimum. Both effects introduce errors in estimating
the original mixing matrix. The upper curve in Fig.~\ref{audstn} shows the averaged Amari 
index over 100 realizations with different noise and mixing matrices.

To reduce this error we modify MILCA to n-MILCA (noisy MILCA). At first we do a 'quasiwhitening' 
with the estimated covariance matrix ${\bf V}=({\bf C}_x-r\bf{1})^{-1/2}$ of the pure 
signals (see, e.g.,~\cite{hyvar2001}, chapter 15) to decorrelate the original sources. As a 
consequence of this, the noise will become now correlated, and with it also the entire 
'quasiwhitened' signal. Because of this we should not minimize $\hat{I}(\phi)$, since in this
way we would introduce a bias as seen in Fig.~\ref{noiseangle} towards wrong values
of $\phi$. Instead we minimize $\hat{I}(\phi)+{1\over 2} \log(1-C_{ij}(\phi)^2)$ where we have
subtracted the `linear' contribution (see Eq.(\ref{MI-ineq})). In Fig.~\ref{audstn} we show 
again the averaged Amari index for the same realizations as used before. Making use of detailed
information on the noise clearly improved the results, except for very small SNR. The amount by 
which it improves depends
on the condition number of the mixing matrix. For matrices far away from singularity (low 
condition number) the quasiwhitening has little effect and there is hardly any difference, while 
for large condition numbers the two mixtures are nearly the same and it is impossible to
obtain good results with either algorithms.

Finally, before leaving this subsection, let us say a few words about outliers. Outliers are just a
special case of noise. Because our MI estimator is based on the k-nearest neighbor distribution, 
outliers make less difficulties Ref.\cite{Learned} than e.g. in kurtosis based algorithms.

\subsection{A Real-World Application}

Finally, let us apply MILCA to a fetal ECG recording from the abdomen and thorax of a pregnant 
woman (8 electrodes, 500 Hz, 5 seconds). We chose this data set because it was analyzed several 
times with different ICA algorithms \cite{Lath,Cardoso98,Meinecke02,alex} and is available on the web 
\cite{ECGdata}.

\begin{figure}
  \begin{center}
    \psfig{file=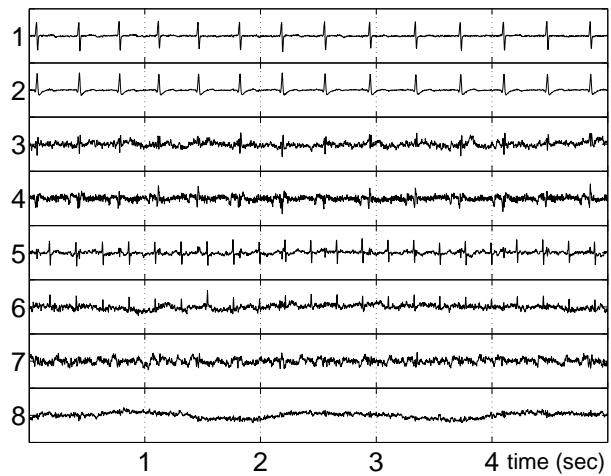,width=8.0cm,angle=0}
  \caption{MILCA output: components after minimizing $I(X_1\ldots X_{8})$ for the heart beat example of 
     Sec.~III.E.}
 \label{ECGMILCA}
 \end{center}
\end{figure}

The output components of MILCA are shown in Fig.~\ref{ECGMILCA} \cite{footnote1}. 
We used $k=30$ neighbors for estimating MI, and to obtain the minima of $\hat{I}_{ij}(\phi)$ we 
fitted with 3 Fourier components.
The success of the decomposition 
is already seen by visual inspection. Obviously, channels 1-2 are dominated by the heartbeat
of the mother, and channel 5 by that of the child. Channels 3, 4, and 6 still contain heartbeat 
components (of mother and child, respectively), but look much more noisy. Channels 7 and 8 seem
to be dominated by noise, but with rather different spectral compositions.

\begin{figure}
  \begin{center}
    \psfig{file=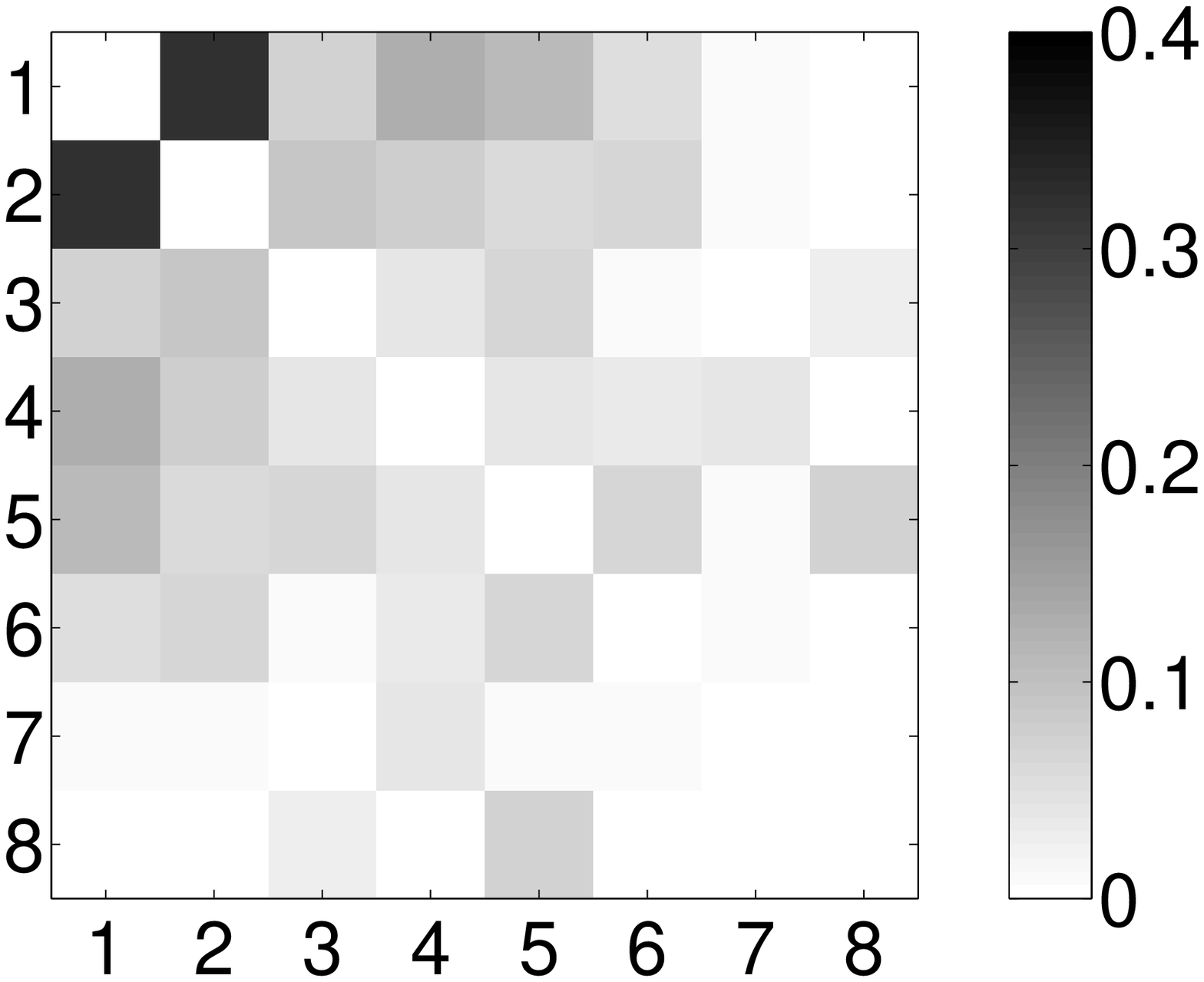,width=3.95cm, height=3.65cm, angle=0}
    \psfig{file=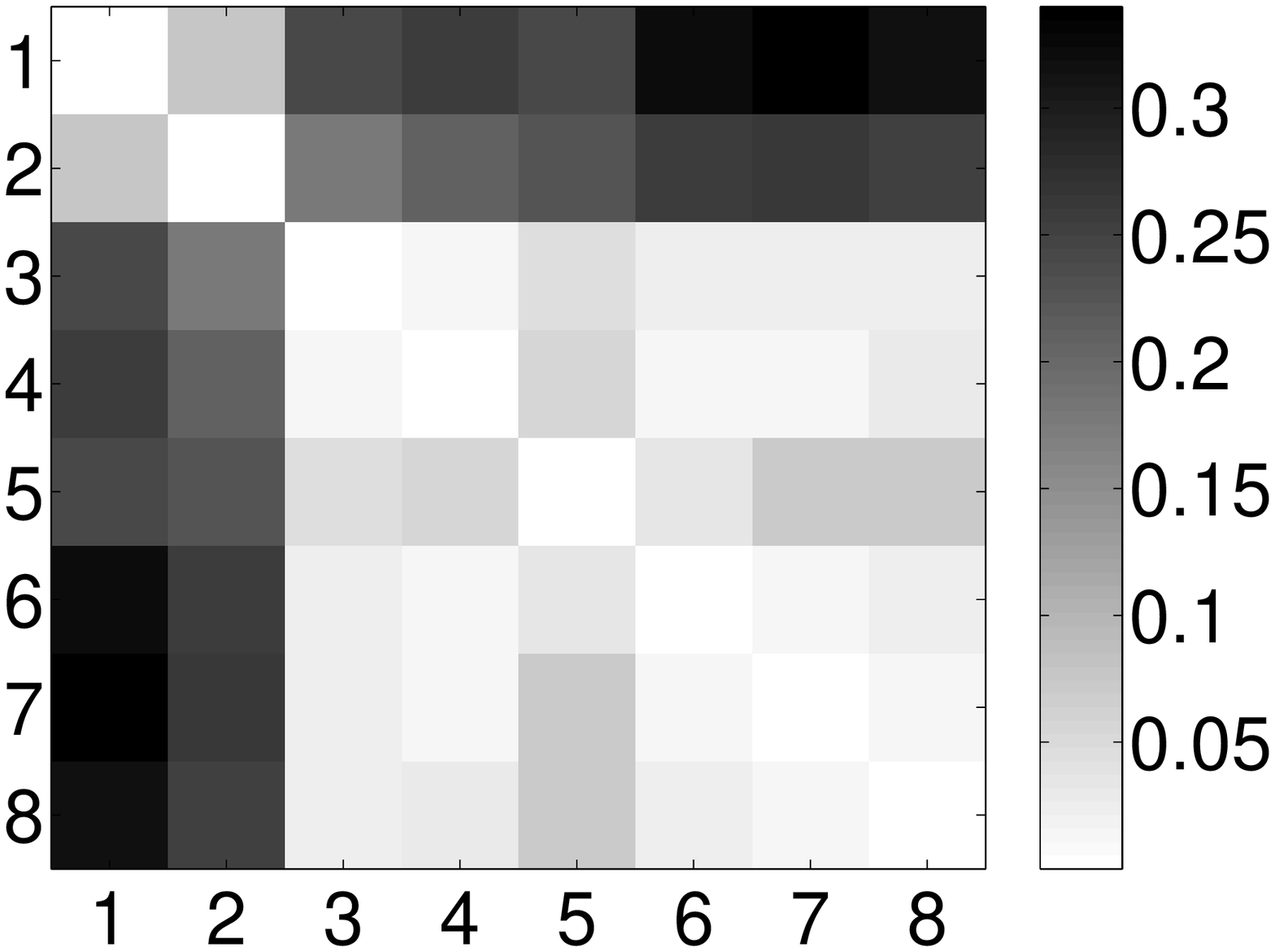,width=3.95cm, height=3.55cm, angle=0}
  \caption{Left panel: $\hat{I}$ between all the pairwise combinations of the signals shown in
     Fig. \ref{ECGMILCA}.  Right panel: Square roots of variabilities $\sigma_{ij}$ of
     $\hat{I}_{ij}(\phi)$. In both panels the values on the diagonal are set to zero.}
 \label{ECGmimatrix}
 \end{center}
 \end{figure}

In order to verify this also formally (which would be essential in any automatic real-time implementation)
we first show in Fig.~\ref{ECGmimatrix} (left panel) the pairwise MIs. We see that most MIs are indeed
small, except the one between the first two components. This indicates again that the first two components
belong to the same source, namely the heart of the mother. But some of the other MIs seem to be 
definitely non-zero, even if they are small. This indicates that the decomposition is not perfect, 
as is also seen by closer inspection of Fig.~\ref{ECGMILCA}.

Finally, we show in the right panel of Fig.~\ref{ECGmimatrix} the variabilities under re-mixing. They
confirm our previous findings. In contrast to the sine/cosine pair in test example (C), the first two
components have non-zero $\sigma$, showing that the distribution in this subspace is not isotropic and 
that one can minimize the interdependence in it by a suitably chosen demixing. Apart from that, the 
biggest values of $\sigma$ are for channels 1, 2, and 5, showing that these channels are most 
reliably and uniquely reconstructed. They are just the channels dominated most strongly by heart beat.

%Figure 9

\section{Cluster Analysis}

We pointed already out that the usual assumption of independent one-dimensional sources as in 
Eq.(\ref{source-x}) is often unrealistic. Take e.g. the ECG discussed in the previous subsection,
and assume that both hearts -- the one of the mother and the one of the fetus -- are independent 
chaotic dynamical systems. A chaotic system with continuous time must have at least 3 excited 
degrees of freedom \cite{ott}. With any generic placement of the electrodes, we should then expect
to pick up $\geq 3$ different components from each heart. These components must be strongly dependent 
on each other, even after having been whitened \cite{broomhead}. Thus each heart must contribute
to at least 3 output components in {\it any} linear ICA scheme. For the mother heart we have 
indeed found 2 components. The fact that we have not clearly identified more dependent components
in the output should be considered as failure of the instantaneous linear algorithm and will be dealt
with more systematically in Sec.~V.

In any case, in view of this we have to expect that outputs in real-world applications are not
independent but come in connected clusters. Moreover, we should expect that even within one cluster
there are more or less strongly connected substructures. We have already discussed in Sec.~III.c a 
simple way how to identify these clusters. In the present section we present a more systematic 
analysis.

\begin{figure}
  \begin{center}
    \psfig{file=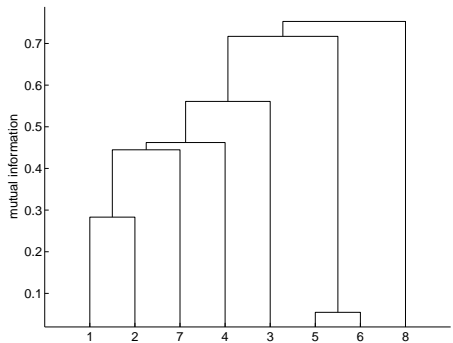,width=6cm}
    \caption{Dendrogram for Fig.~\ref{ECGMILCA}. The height of each cluster $(ij)$ corresponds to
     $\hat{I}(X_i,X_j)$ ($k$=6).}
    \label{ClustECG}
\end{center}
\end{figure}

Our strategy is to estimate a proximity matrix from the MIs, and then to use a hierarchical clustering 
algorithm to obtain a dendrogram. No thresholds are used in constructing the dendrogram, i.e. it is 
constructed without making any decision about which MILCA output channels are independent or not. Only 
after its construction we decide, usually on heuristic reasons and based on arguments of practicality
and usefulness, which channels are actually grouped together. This is more convenient, usually, than
the algorithms of \cite{Bach,phinv,topica} where this decision stands at the starting point of the 
algorithm or is an essential part of it.

A first technical problem concerns the choice of the proximity matrix. One might
be tempted to use MI directly. But we want to include the possibility that some of the channels to 
be grouped together are already multidimensional by themselves. In this case, using MI would introduce
a bias: multivariate channels not only tend to carry more information than univariate ones, they 
also will have larger MIs. Therefore we propose to use as a similarity measure \cite{alex2}
\be
  P_{ij} =  {\hat{I}(\hat{s}_i,\hat{s}_j) \over \dim(\hat{s}_i)+\dim(\hat{s}_j)},       \label{d}
\ee
where $\dim(x)$ is the dimension of the variable $x$, i.e. the number of its components.

In most cluster algorithms the proximity matrix P is used only for the first step. In the subsequent
steps, proximities for clusters are derived from it in some recursive way \cite{jain-dubes}. In 
the present paper we propose to use `MI-based Clustering' (MIC) \cite{alex2} which is based on 
the grouping property Eq.(\ref{grouping}). Thus, a cluster of output channels is just characterized 
by the multivariate signal formed by the tuple of its individual channels, and the proximity 
measure is still given precisely by Eq.(\ref{d}) at each level of the hierarchy.

In summary, our cluster algorithm is as follows. We start with $n$ (usually univariate) MILCA output 
channels $\hat{s}_i\;,\;\;i=1,\ldots n$, and we compute $P_{ij}$ according to Eq.(\ref{d}). After
that, we enter the following recursion:

\begin{enumerate}
 \item Find the pair with minimum distance in the matrix, say clusters $i$ and $j$;
 \item Combine the clusters $i$ and $j$ to a new cluster $(ij)$ with multivariate data $\hat{s}_{ij}$,
    and attribute to it a height $\hat{I}(\hat{s}_i,\hat{s}_j)$ in the dendrogram. Thereby the total 
    number of clusters is reduced by one, $n\leftarrow n-1$;
 \item If the new value of $n$ is 1 then exit; else
 \item update the proximity matrix $P_{ij}$ and go to 1.
\end{enumerate}

The dendrogram obtained in this way for the ECG data of Sec.~III.E is shown in Fig.~\ref{ClustECG}.
In this figure two clusters are clearly distinguishable, the mother cluster containing channels
(1, 2, 3, 4, 7) and the fetus cluster formed by channels 5 and 6. This agrees perfectly with the 
interpretation given in Sec.~III.E. One can of course debate whether, e.g., channel 7 belongs to the 
mother cluster or not, but this can be decided as it seems most convenient, and it will in 
general have little effect on any conclusions. One way to make use of such a clustering is in
cleaning the data and separating the individual sources. For that, one prunes everything except
the wanted cluster, and reconstructs the original channels by applying the inverse of the matrix
$\bf W$. Results obtained in this way will be shown in the next section, after having discussed
how to take into account temporal structures.

\section{Using Temporal Structures}

\subsection{Instantaneous Demixing that Minimizes Delayed Mutual Informations}

Until now we have not used any time structure in the signals. In the following we shall assume
the signals to be stationary with finite autocorrelation times. ICA-algorithms in the literature
either use no time information at all (JADE\cite{Cardoso93}, FastICA\cite{hyvar2001},
INFOMAX\cite{bell}, ...) or, if they do use it, they use only second order statistics 
(AMUSE~\cite{tong}, TDSEP~\cite{Meinecke02},...). The first group is not able to decompose two 
Gaussian signals with different spectra, while the second group is not able to separate two
temporally white signals with different amplitude distributions. Obviously one has to make 
use of time structure {\it and} higher order statistics, to obtain optimal results in general
\cite{mueller,Amaribook}. This is precisely what we will do in this subsection.

\begin{figure}
  \begin{center}
    \psfig{file=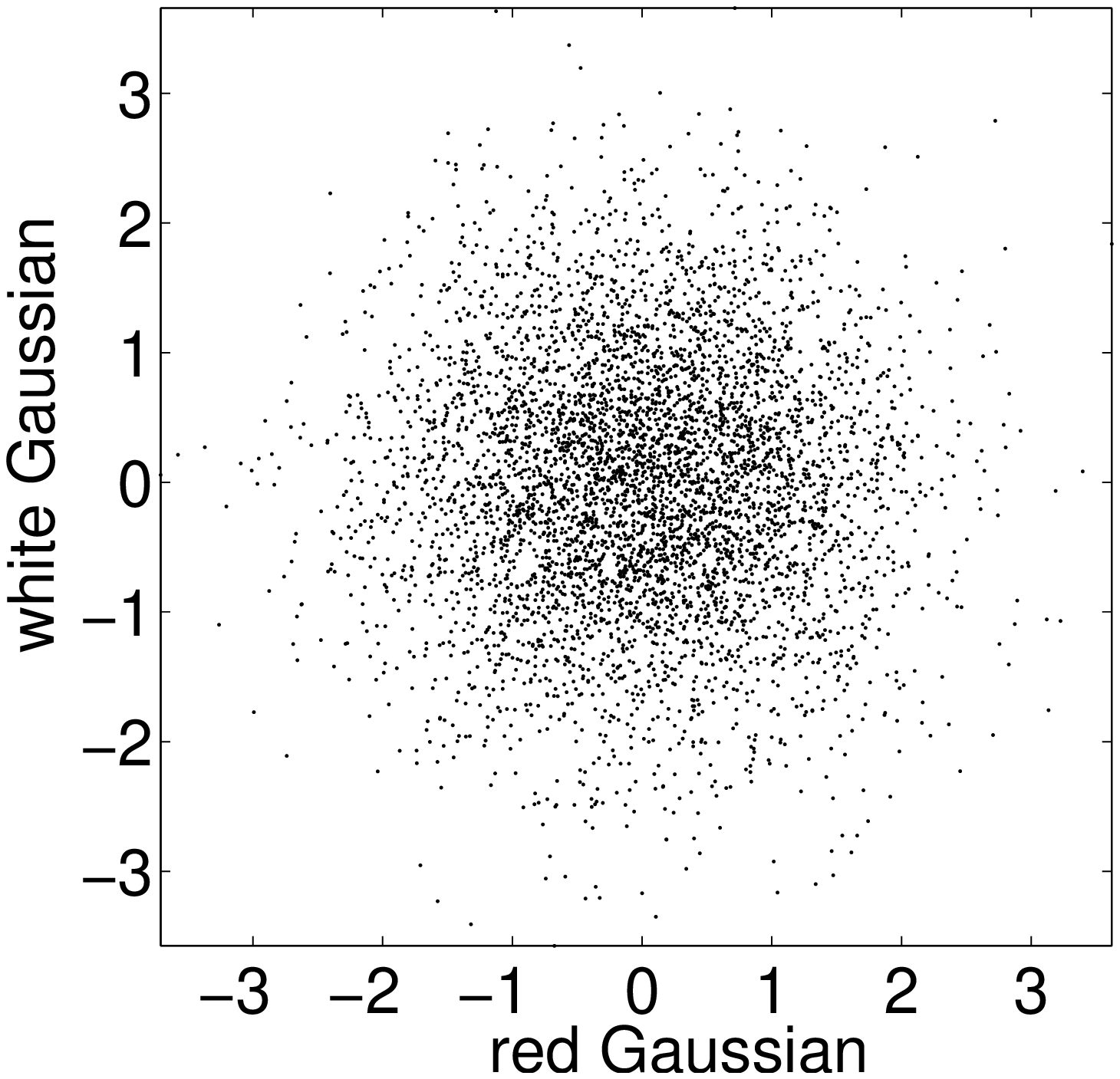,width=3.8cm, height=3.5cm,angle=0}
    \psfig{file=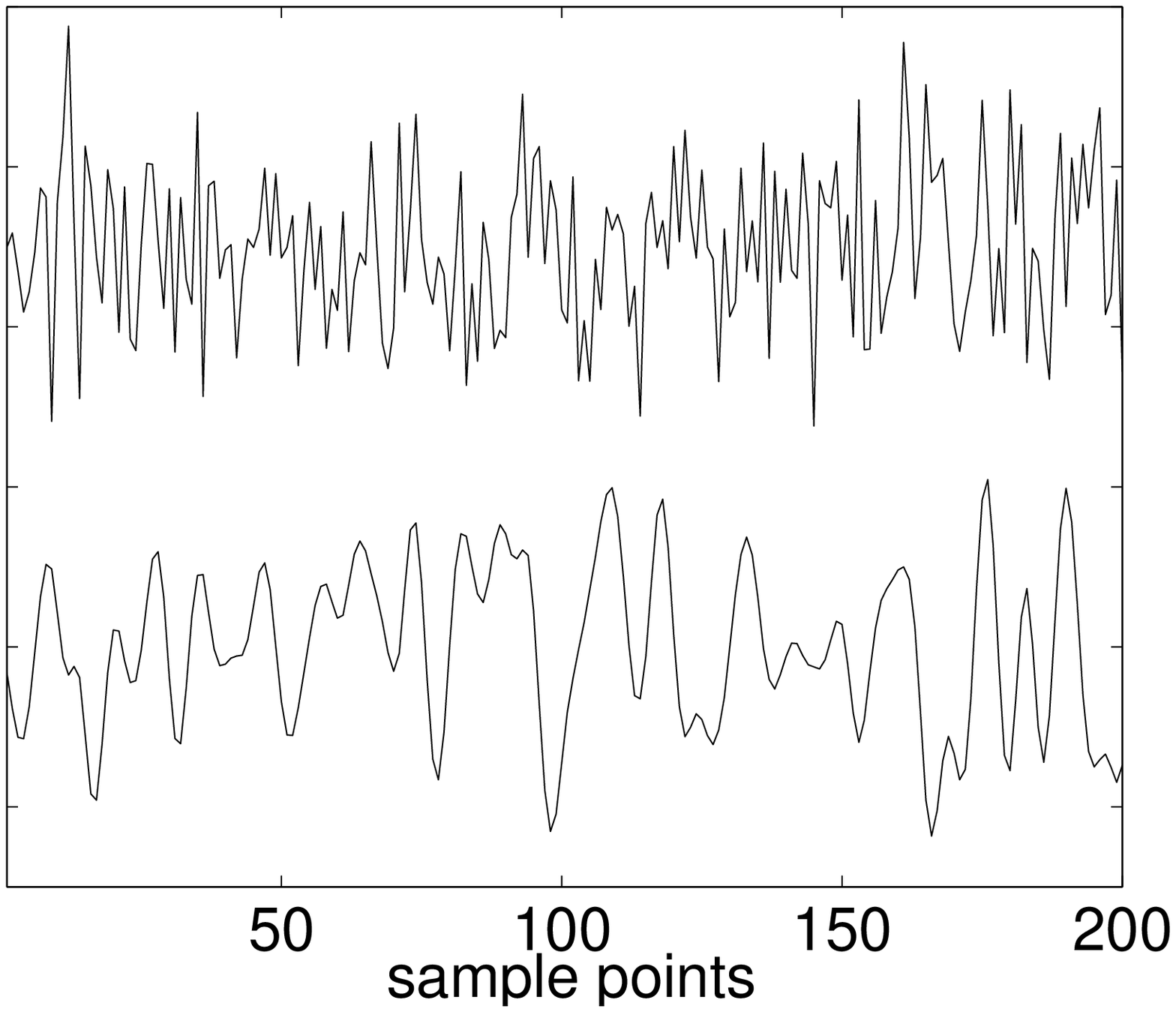,width=3.8cm, height=3.5cm,angle=0}
%   \vskip 0.3 true cm
 \caption{Left panel: scatter plot of the two Gaussian sources with different spectra.
Right panel: Output of the modified MILCA algorithm ($\tau=1$ and $m=2$), where white Gaussian 
is on top and red Gaussian is on bottom.}

 \label{redGauss}
 \end{center}
\end{figure}

\begin{figure}
  \begin{center}
    \psfig{file=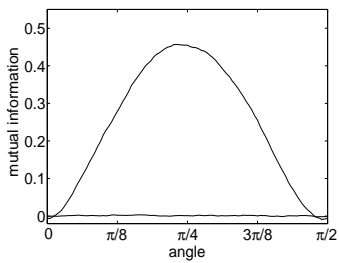,width=4.5cm,height=3.4 cm,angle=0}
%   \vskip 0.3 true cm
 \caption{Change of $\hat{I}$ under rotation, for the Gaussian model shown in Fig.~\ref{redGauss}.
   The nearly horizontal curve shows the behavior without, the sinusoidal one the result with
   using delay embedding. Here the actual mixing angle is $0$.}
 \label{redGauss2}
 \end{center}
\end{figure}

\begin{figure}
  \begin{center}
    \psfig{file=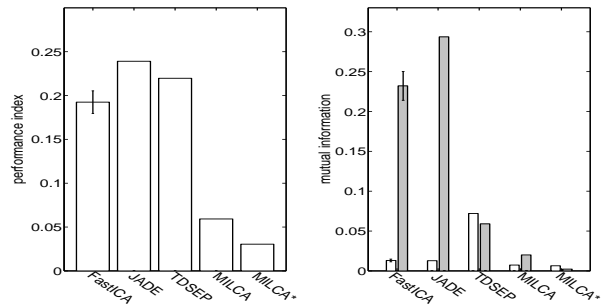,width=8cm,height=4 cm,angle=0}
 \caption{
    Test problem (B) of Sec.III, consisting of 5 input channels (compare with Fig.\ref{algocomp}).
   Algorithm ``MILCA$^*$'' now refers to the minimization of Eq.(\ref{MIembed}). The gray bars
   on the right panel show the full MI given in Eq.(\ref{MIembed}). The embedding parameters are $m=2$, $\tau=1$.}
 \label{perfdelay}
 \end{center}
\end{figure}

Normally the first step in nonlinear time series analysis of univariate signals is delay embedding
\cite{Kantz}: One constructs a formally $m$-variate signal, for any $m>1$, by simply forming
$m$-dimensional `delay vectors' with a suitably chosen delay $\tau$,
\be
  {\bf x}(t) = [x(t-\tau),x(t-2\tau),\ldots x(t-m\tau)]^T.           \label{delay}
\ee
Thus one characterizes the ``state" of a signal at time $t$ by giving not its value at
$t$ itself, but at $m$ previous times. This makes of course sense only when there is any time
structure in the signal. Similarly we can also embed multivariate signals. For $n$ measured
channels one obtains thereby an $n\times m$ `delay matrix'
\be
  {\bf X}(t) = [{\bf x}_1(t), \ldots {\bf x}_n(t)].
\ee

To decompose an instantaneous linear mixture of $n$ signals with either non-Gaussian statistics
or with non-trivial time structure, we propose to simply minimize the MI,
\be
   \hat{I}({\bf s}_1(t), \ldots {\bf s}_n(t)) \stackrel{!}{=} \min.      \label{MIembed}
\ee
Notice that we have here considered the delay vectors as joint entities, i.e. we
do not include in Eq.(\ref{MIembed}) the MIs between the different delays of the same $x_i$.
More explicitly \cite{footnote4},
\bea
   I({\bf x}_1(t), \ldots {\bf x}_n(t)) & = & I(x_1(t-\tau),\ldots x_1(t-m\tau),\nonumber\\
                                        &\; & x_2(t-\tau),\ldots x_2(t-m\tau),\ldots \nonumber\\
                                        &\; &x_n(t-\tau),\ldots x_n(t-m\tau))          \\
    &-& \sum_{i=1}^n I(x_i(t-\tau),\ldots x_i(t-m\tau))                       \nonumber \\
    & = & \sum_{i=1}^n H({\bf x}_i(t)) - H({\bf x}_1(t), \ldots {\bf x}_n(t)) \nonumber
        \label{Idelay}
\eea
To minimize this, we proceed again as in Sec.III, i.e. we decompose the rotation needed to
minimize $\hat{I}({\bf x}_1(t), \ldots {\bf x}_n(t))$ into rotations within each of the
$n(n-1)/2$ coordinate planes. Each of the latter rotations still involves rotations of $m$
delay coordinate pairs, but this can be further decomposed into $m$ rotations where only
one delay coordinate pair is rotated. We thereby obtain
\bea
   & &I(\ldots {\bf x}_i'(t), \ldots {\bf x}_j'(t)\ldots) -
      I(\ldots{\bf x}_i(t), \ldots {\bf x}_j(t)\ldots) \nonumber \\
   & & = I({\bf x}_i'(t), {\bf x}_j'(t)) - I({\bf x}_i(t), {\bf x}_j(t)) \nonumber \\
   & & = \; I(x_i(t-\tau),\ldots x_i(t-m\tau)) \nonumber \\
   & & \;\; + I(x_j(t-\tau),\ldots x_j(t-m\tau)) \nonumber \\
   & & \;\; - I(x'_i(t-\tau),\ldots x'_i(t-m\tau)) \nonumber \\
   & & \;\; - I(x'_j(t-\tau),\ldots x'_j(t-m\tau)) \nonumber \\
   & & \;\; + m\;[ I(x'_i(t),x'_j(t)) - I(x_i(t),x_j(t))] \;,
\eea
where we have used in the last term the fact that $I(x'_i(t),x'_j(t))$ is independent of $t$
due to stationarity. If $m=2$, this is again a sum of pairwise MIs. If $m>2$, we have to
estimate $m$-dimensional MIs directly.

To illustrate this on a simple example, let us assume two channels where $x_1(t)$ and
$x_2(t)$ are instantaneous mixtures of two Gaussian signals with the same amplitude distribution but
with different spectra: $x_1$ is white (iid), while $x_2$ is red and was obtained by filtering
with a Butterworth filter of order 6 and with cutoff frequency 0.3.
For simplicity we assume the mixing to be a pure rotation. Then a scatter plot of
the vectors $(x_1(t),x_2(t))$ is completely featureless, see Fig.~\ref{redGauss}(left),
and will not allow a unique decomposition. But using delay embedding
with $m=2$ is sufficient to obtain the original sources (Fig.~\ref{redGauss}, right panel).

\begin{figure}
  \begin{center}
    \psfig{file=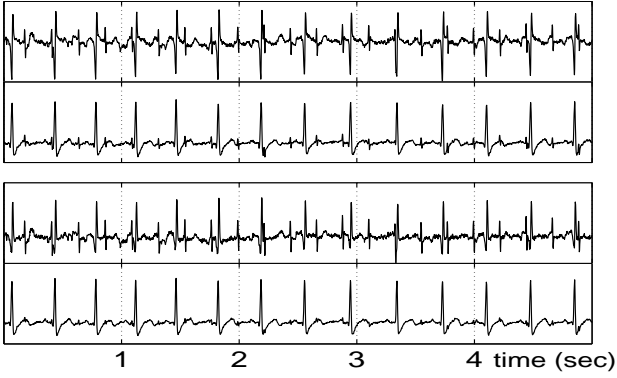,width=8.3cm,height=5cm,angle=0}
 \caption{Upper panel: Two channels of the ECG of a pregnant woman.
          Lower panel: MILCA output from these two channels.}
 \label{ECG2}
 \end{center}
\end{figure}

\begin{figure}
  \begin{center}
    \psfig{file=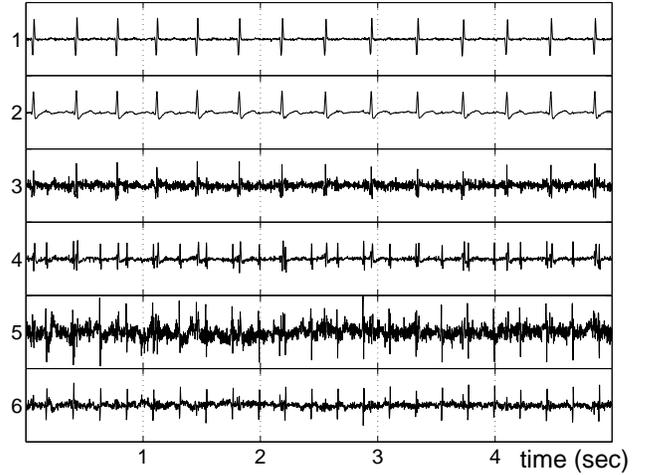,width=8.3cm,angle=0}
 \caption{MILCA output from the delay embedded two channel ECG with embedding dimension $m=3$.}
 \label{ECG2embICA}
 \end{center}
\end{figure}

Similarly good results were obtained with the less trivial examples of previous sections. In particular, we
tested the algorithm on test problem (B) of Sec.~III.B (Fig.~\ref{perfdelay}). The performance of MILCA is
improved substantially, even with $m=2$. The delayed MI (Eq.(\ref{MIembed})) which make use of the time structure
serves as a better performance value (Fig.~\ref{perfdelay} (right)). Now JADE and FastICA are also heavily punished for not
separating one audio signal from Gaussian noise (as one can see the MI for TDSEP is nearly unchanged
because the time correlation in the output is minimal).

\subsection{Demixing with Delays}

The most general linear demixing ansatz for a stationary system assumes superpositions 
of the observed signals {\it with delays}. Using up to $m$ delays $\tau, 2\tau, \ldots m\tau$, 
we thus make the ansatz (see e.g. Ref.~\cite{hyvar2001}, chapter 19)
\bea
   \hat{s}_i(t) &=& \sum_{j=1}^N \sum_{k=1}^m w_{ij}^k\;x_j(t-k\tau) \nonumber \\
                &=& \sum_{j=1}^N {\bf w}_{ij}{\bf x}_j(t) \;,
   \label{filter}
\eea             
where ${\bf x}_j(t)$ is a delay vector as defined in Eq.(\ref{delay}) and 
\be
  {\bf w}_{ij} = [w_{ij}^1\ldots w_{ij}^k].           \label{w-delay}
\ee
Since we have now linear superpositions of $n\times m$ measurements $x_j(t-k\tau)$ on the 
right hand side, we can also determine the same number of $\hat{s}_i(t)$ for each value of 
$t$, i.e. the index $i$ in Eq.(\ref{filter}) runs from 1 to $nm$.

This ansatz is obviously more appropriate than instantaneous mixing, if the signals $x_i(t)$ are 
themselves superpositions of delayed sources. If they involve a finite number of delays,
\be
   x_i(t) = \sum_j \sum_{k=1}^{m'} a_{ij}^k s_j(t-k\tau),    \label{mix}
\ee
Eq.(\ref{filter}) with finite $m$ would {\it not} give the {\it exact} demixing, since inverting 
Eq.(\ref{mix}) would require an infinite number of delay terms. Also, Eq.(\ref{filter}) 
in general does not correspond to the inverse of Eq.(\ref{mix}), because its solutions are in
general not components of any delay vectors. But it should definitely be a 
better ansatz than the instantaneous Eq.(\ref{source-x}). 

Apart from that, we would anyhow not expect Eq.(\ref{mix}) to be the correct model in 
most applications. The main reason why we believe that Eq.(\ref{filter}) is useful in many 
applications is that it can cope much better with the situation discussed at the beginning of 
Sec.~IV. Assume for the moment that there is a single source. Different sensors (as, e.g., 
different ECG contacts) typically see different projections of this source, and the signals 
$x_i(t)$ can therefore be considered as different coordinates 
describing its dynamics. As pointed out by Takens \cite{Kantz}, delayed values of one single 
signal can also be considered as different coordinates. Our demixing ansatz basically reflects 
the hope that suitable superpositions of delayed values of $x_i$, say, can mimic any other 
signal $x_j$.

To illustrate this, we consider again the above ECG recording. We assume for the moment that 
only the two channels with the most pronounced fetus heartbeat are available and try to 
decompose them into mother and fetus heartbeat. These two channels are shown in 
Fig.~\ref{ECG2} (top).  They are still dominated by the mother heartbeat. But the R peak of 
the mother has a very different shape in both channels: In the lower trace it is mainly 
positive, while it has both positive and negative components in the upper. It is therefore 
clear that there cannot exist an instantaneous superposition to which the mother's heartbeat 
does not contribute. Instantaneous ICA {\it must} fail for this case, as is indeed seen in 
the lower two traces of Fig.~\ref{ECG2}.

In order to obtain the least dependent components obtainable with Eq.(\ref{filter}), we minimize 
again the MI. But now, in contrast to the previous subsection, the output variable $s_i(t)$ are 
not delay coordinates of any sources, and therefore we must minimize the full MI between {\it all}
$s_i(t)$,
\be
   \hat{I}(s_1(t),\ldots s_{nm}(t)) \stackrel{!}{=} \min.      \label{MIembed-delay} 
\ee
The minimization is done again, as in all previous cases, by performing successive transformations
in 2-dimensional subspaces and by using Eq.(\ref{dI}). In terms of the actual algorithm, the 
only difference to the previous subsection is that we now make rotations in {\it all} subspaces.

\begin{figure}
  \begin{center}
    \psfig{file=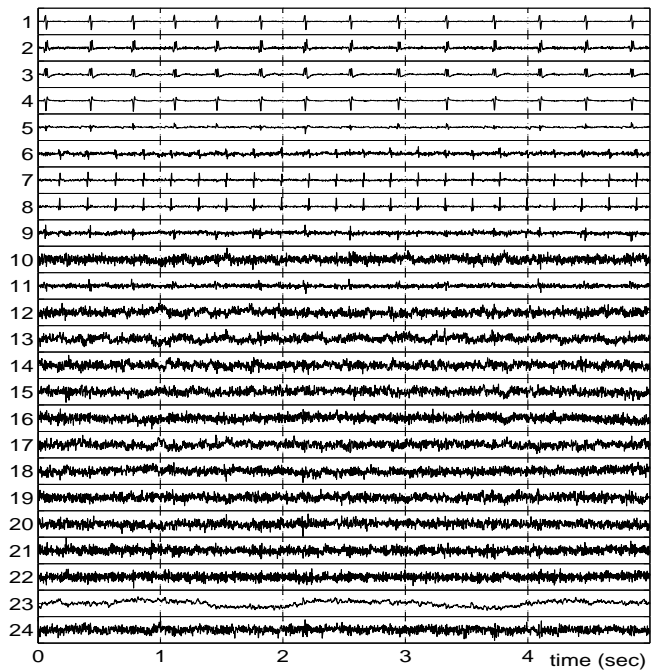,width=8.6cm,height=8.9cm,angle=0}
%   \vskip 0.3 true cm
 \caption{MILCA output from the embedded eight channel ECG (k=100, m=3)}
 \label{ECG24embICA}
 \end{center}
\end{figure}

\begin{figure}
  \begin{center}
    \psfig{file=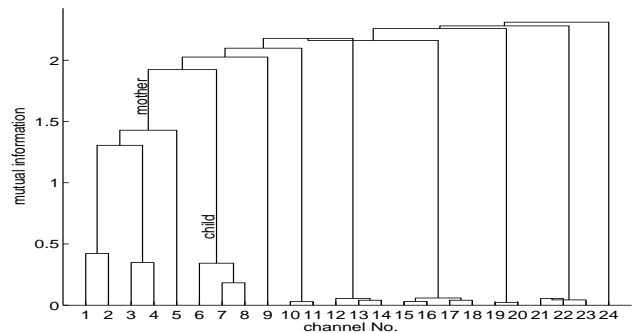,width=8.4cm,height=4.4cm,angle=0}
%   \vskip 0.3 true cm
 \caption{Dendrogram for Fig.~\ref{ECG24embICA}. Heights of each cluster correspond to $I(X_i,X_j)$ 
      of the cluster $ij$ ($k$=3).}
 \label{clustECG24}
 \end{center}
\end{figure}

In our application to the fetal ECG we use embedding dimension $m=3$ and the smallest possible 
delay, $\tau = 1/500\; {\rm s}^{-1}$. Results for the two channels shown in Fig.~\ref{ECG2} are 
now shown in Fig.~\ref{ECG2embICA}. The separation is now improved. Although we still have one 
output channel where mother and fetus are strongly mixed (channel \#4), channel \#6 is now practically
pure fetal heartbeat.

Finally, we applied this method to all 8 channels of the ECG. Using again $m=3$ gives altogether 
24 output channels. They are shown in (Fig.~\ref{ECG24embICA}), and we can clearly see which ones 
are dominated by the mother heartbeat, which by the fetus, and which by noise. In order to do 
this more objectively, we again apply the cluster algorithm of Sec.~IV, with the result shown 
in Fig.~\ref{clustECG24}. There one can clearly see two big clusters corresponding to the mother
and to the fetus. There are also some small clusters which should be considered as noise.

For any two clusters (tuples) $X=X_1\ldots X_p$ and $Y=Y_1\ldots Y_q$ one has 
$I(X_1,\ldots Y_q) \geq I(X)+I(Y)$. This guarantees, if the MI is estimated
correctly, that the tree is drawn properly, i.e. each parent node is above the two daughter 
nods. The two slight glitches (when clusters (1 - 14) and (15 - 18)
join, and when (21 -22) is joined with 23) result from small errors in estimating MI. They
do not affect our conclusions.

\begin{figure}
  \begin{center}
    \psfig{file=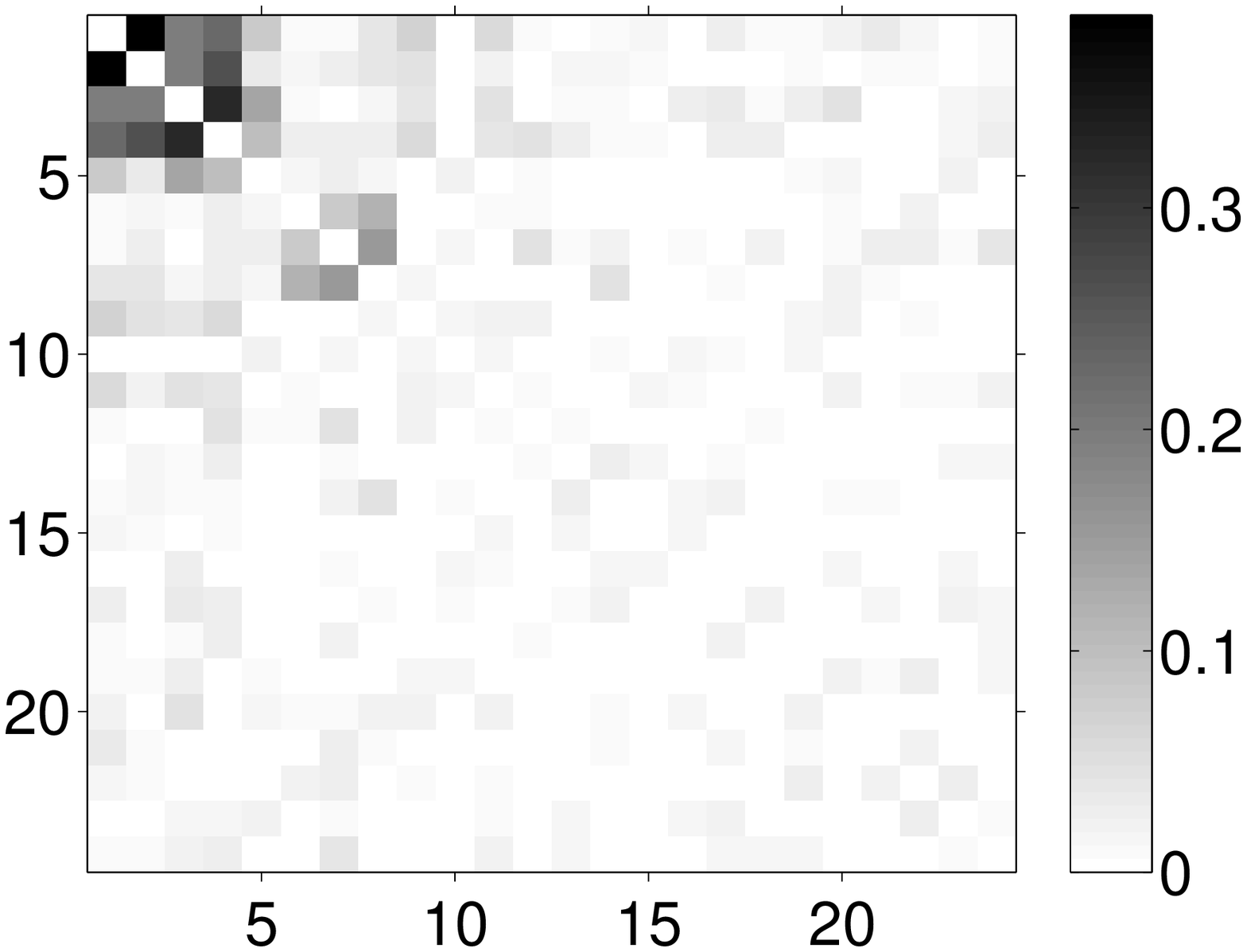,width=4.1cm,angle=0}
    \psfig{file=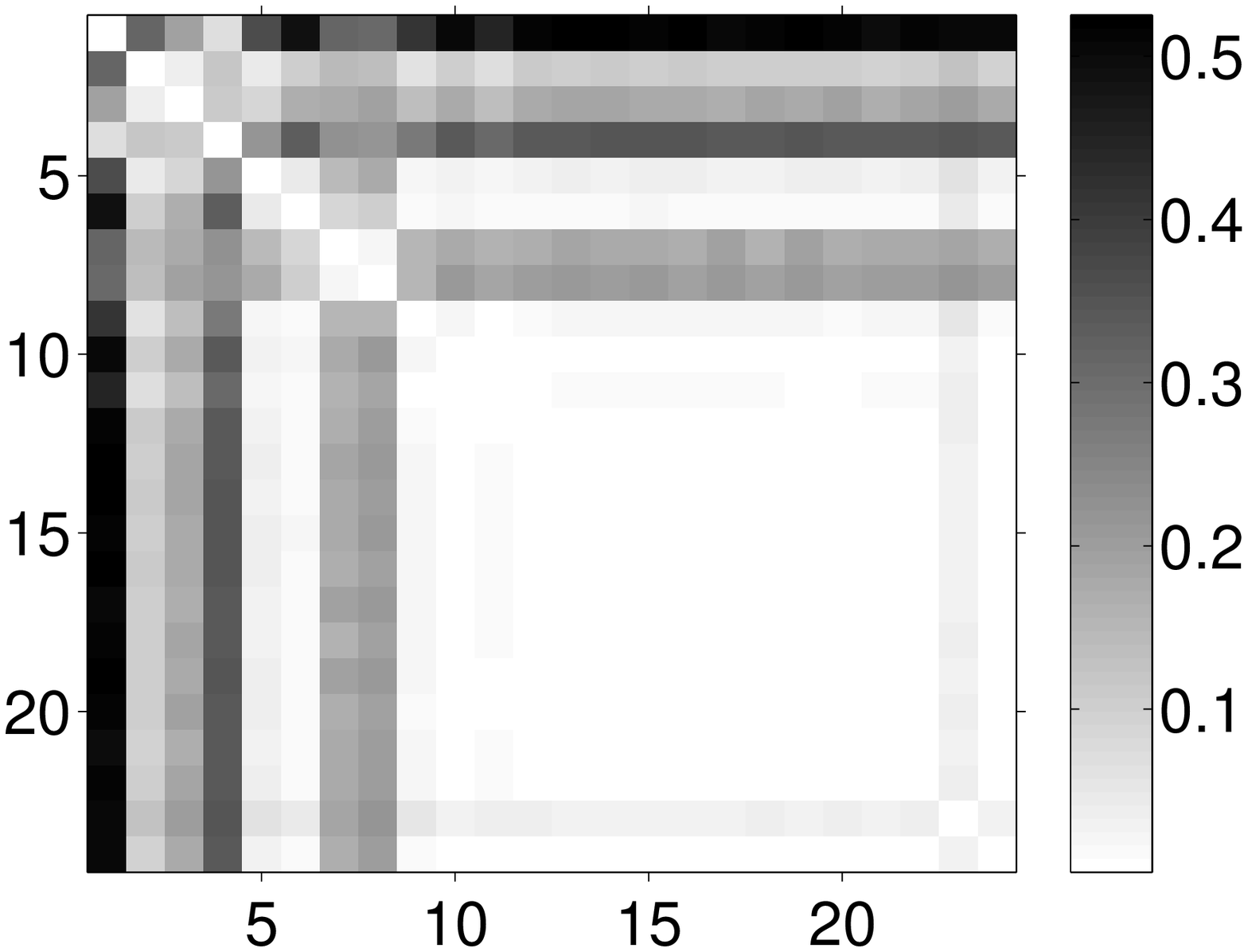,width=4.1cm,angle=0}
  \caption{ Left panel: Pairwise MIs between the estimated components shown in
     Fig.~\ref{ECG24embICA}. Right: Square roots of variabilities ${\sigma_{ij}}$ of
     $I(X_i, X_j)$ (with $k=6$). Elements on the diagonal have been set to zero.}
 \label{matrixECG24}
 \end{center}
\end{figure}

In Fig.~\ref{matrixECG24} we show the matrices of pairwise MIs (left panel) and of pairwise 
variabilities (right). They are as expected, and they show much more pronounced structures 
than the matrices without delay embedding (Fig.~\ref{ECGmimatrix}). For the MIs one can see a 
clear block structure, i.e. the mother and fetus components are now indeed more independent, 
as suggested also from the traces themselves. From the right panel we see that the main mother
channels (1-4) and the fetus channels (7-8) are very stable. The rest is mostly noise, and 
is not stable as indicated by the very small variabilities.

\begin{figure}
  \begin{center}
    \psfig{file=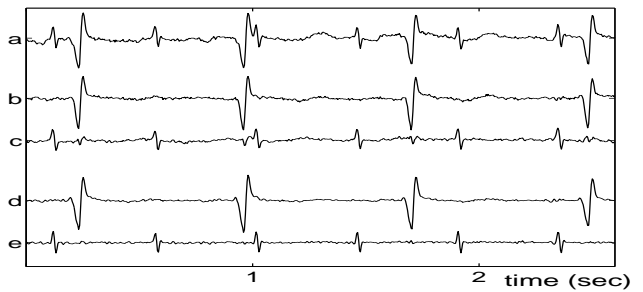,width=8.4cm,height=3.8cm,angle=0}
%   \vskip 0.3 true cm
 \caption{Short segment from the original ECG (a), of the mother and fetus contributions estimated
     without delay embedding (b,c), and of the two contributions estimated with delay embedding (d,e).}
 \label{ECGzoom}
 \end{center}
\end{figure}

The final result of MILCA is obtained by pruning everything not belonging to the cluster of 
interest,
\be
  \hat{s}_i(t) \to {\cal P}_C\;\hat{s}_i(t) \equiv \left\{ \begin{array}{rl}
             \hat{s}_i(t) &\qquad  i \in {\rm cluster}\;\; C \\
                    0     &\qquad  {\rm else} \end{array} \right.
\ee
and performing the back transformation. At this stage there arises the problem that the 
reconstructed signals 
\be
   \hat{x}_{j,k}(t;C) = {\bf W}^{-1}_{(j,k),i} {\cal P}_C\;\hat{s}_i(t) \;, \quad 
    {\bf W}_{i,(j,k)} = w^k_{ij}
\ee
are in general not delay vectors, i.e.
\be
   \hat{x}_{j,k+1}(t;C) \neq \hat{x}_{j,k}(t-\tau;C)\;.
\ee
In view of this, one has to make some heuristic decision what to use as a cleaned signal. We
use simple averages,
\be
   \hat{x}_j(t;C) = {1\over m} \sum_{k=1}^m \hat{x}_{j,k}(t+k\tau;C)\;.
\ee
We do not show all 8 full traces for the mother and fetus, because this would not be very 
informative: The results are too clean to be judged on this scale. Instead we show 
in Fig.~\ref{ECGzoom} blow-ups of one of the original traces and the contributions to it from 
the mother and from the fetus. The separation is practically perfect.

Before leaving this section, we should point out that one can, in principle, also construct
algorithms in between those of the last two subsections. In subsection A we had used delays 
to minimize the lagged MI, but we had not used the delays in the demixing. In the present 
subsection, we have used the same delays both for minimizing MI and for demixing. A generalization
consists in using $m$ delays in the demixing, but minimizing the MI with additional $m'$ delays.
Thus we make the same demixing ansatz Eq.(\ref{filter}) as above, but we minimize 
\be
   \hat{I}({\bf s}_1(t), \ldots {\bf s}_{nm}(t)) \stackrel{!}{=} \min.
\ee
where we have used the definition of $I({\bf s}_1(t), \ldots)$ given in Eq.(\ref{Idelay}), and 
$\hat{\bf s}_i(t) = [\hat{\bf s}_i(t-\tau),\hat{\bf s}_i(t-2\tau),\ldots \hat{\bf s}_i(t-m'\tau)]^T$.
Up to now, we have not yet applied this to any problem.

\section{Discussion}

There is by now a huge literature on independent component analysis. Therefore, most of our 
treatment is related in some form to previous work. One of our basic premises was that we did 
not so much care about speed, but we wanted as precise a dependency measure as possible. Our 
claim that this is provided in principle by MI is of course not new. But we believe that our 
estimator via $k$-nearest neighbor statistics is new and provides the most precise mutual 
information estimate.  It is closely related to similar estimators for {\it differential 
entropies} which had been used in \cite{Pham,Learned}, and the quality of our results in the 
most simple 2-d blind source separation problem is very similar to the one in \cite{Learned}. 
The main virtue of our MI estimator, compared to all previous MI estimators, is the numerical 
fact that it becomes unbiased when the two distributions are independent. 

While using differential entropies instead of MI would give the same quality and somewhat simpler
codes for the basic blind separation problem, using MI has other advantages: with it we can estimate
the residual dependencies between the output components. Our use of this knowledge for estimating
the output uniqueness and robustness, by measuring how the dependencies change under re-mixing, 
seems to be new. Previous authors used for this problem resamplings and/or noise addition 
\cite{Meinecke02,noiseinj,icasso}.

In addition to this, we used the MIs between the outputs to cluster them, and we then
used this clustering to obtain the contributions of the individual (multidimensional) sources 
to the measured signals. The observation that ``independent" component analysis will in general, 
when applied to real world data, not give independent components is not new either 
\cite{Bach,topica,Cardoso98}. We stress it by calling our 
approach a ``least dependent" component analysis. Our detailed implementation of this idea seems 
to be new, not the least because our clustering algorithm is novel and uses a specific property 
of MI not shared by other contrast functions.

Although the extension of our algorithm to data with time structure discussed in Sec.V.A seems
straightforward, this strategy of combining in the contrast function deviations from Gaussianity
both at equal times and at non-equal times has been considered in very few papers only \cite{mueller,Amaribook}.
The present paper is the first which uses directly MI for combining these two aspects. In Sec. V. was
shown that this can substantially improve the separation e.g. of audio signals.

Both the ansatz of Sec.V.A and the method of demixing with delays in Sec.V.B are entirely based
on MI, and use essentially the same algorithm. Therefore, also the generalization mentioned at 
the end of Sec.V.B uses essentially the same basic algorithm. This last generalization was 
never considered before, but demixing with delays is of course a very widely treated concept 
(see e.g. \cite{hyvar2001}). It is usually called `convolutive mixing'.
In our presentation we stressed several features which are typically overlooked. One is that 
the `convolutive' {\it demixing} ansatz Eq.(\ref{filter}) is in general, when the sources 
$s_i(t)$ are not strictly independent, {\it not} equivalent to a convolutive {\it mixing} 
ansatz, because the sources then will not be components of delay vectors. This is also the 
reason why we avoided the term `convolutive mixing'.

Just as ICA may be considered as a generalization of principle component analysis (PCA) to 
non-Gaussian contrast functions, mixing with delays is a generalization of multivariate 
{\it singular source analysis} (SSA) \cite{kepenne,ghil} to include non-Gaussianity. 
Univariate SSA, see e.g. \cite{broomhead,vautard},
is often considered as an alternative to Fourier decomposition and has found many applications, 
while multivariate SSA was mainly used in geophysics. Indeed, we consider blind source separation
algorithms based on temporal second order statistics (AMUSE, TDSEP) as more closely related to 
multivariate SSA than to other ICA methods based on nonlinear contrast functions.

While we discussed also a number of other applications and test models, our main test problem 
was the ECG of a pregnant woman, and the task was mainly to extract a clean fetal ECG. We have 
chosen this partly because this ECG was already used in previous ICA analyses 
\cite{Lath,Cardoso98,Meinecke02}. We believe that our method clearly outperformed these and
gives nearly perfect results, although we should admit that the signals to start with were 
already exceptionally clean. It would be of interest to see how our method performs on 
more noisy (and thus more typical) ECGs.
Obtaining fetal ECGs should be of considerable clinical interest, although it is not 
practiced at presence, mainly because of the formidable difficulties to extract them with 
previous methods. In this respect we should mention the seminal work of \cite{kaplan,richter}
where fetal ECGs were extracted even from {\it univariate} signals using locally nonlinear
methods. It would be interesting to see how our method compares with such a nonlinear 
method when the latter is used for multivariate signals.

Throughout the paper, we used {\it total} MI as a contrast function. One might a priori 
think that the sum of all pairwise MIs would be easier to estimate, and could be as useful
as the total MI. Neither is true. One reason for the efficiency of our algorithm is that
{\it changes} of the total MI under linear remixings can be estimated by computing only 
pairwise MIs (except for the method of Sec.~V.A with embedding dimension $m>2$). 
Thus one needs to compute the full high-dimensional MI only once. For all changes during 
the minimization, computing pairwise MIs is sufficient. But this does not mean that 
total MI is essentially a sum of pairwise MIs. We showed in the appendix that this can be 
very wrong. And we found in more realistic applications that the sum over all pairwise MIs 
sometimes {\it increases} when we minimize total MI. Therefore we consider the sum over
all pairwise MIs as a very bad contrast function. 

This is somewhat surprising if one considers ICA as a generalization of PCA. PCA can be viewed 
as minimalization of the sum over all squared pairwise covariances. But we believe that 
this close relation between ICA and PCA is somewhat misleading anyhow. It is usually 
based on this analogy that the data are first {\it pre-whitened}, before the ICA analysis
proper is made, which is then restricted to pure rotations. We showed by means of a counter
example that this can lead to a solution which does {\it not} have minimal MI. This was 
a rather artificial example, and the problem might not be serious in practice (all our 
results were obtained, for simplicity, with prewhitening). But one should keep it in 
mind in future applications.

Finally we should point out that Eqs.~(\ref{grouping},\ref{dI}) hold for the exact MI, but are 
only approximately true for our estimators. Therefore, working directly on higher dimensional 
MIs, without breaking their changes down to 2-dimensional contributions, can give slightly 
different results. We found no big systematic trends, although we expect in general that 
estimates using the smallest dimensions are most reliable. The reason is that they are based 
on smaller distances for fixed $k$, or use larger $k$ when using the same distances. The first 
reduces systematic errors, the second statistical ones. The decrease in CPU time when 
using Eq.~(\ref{dI}) to decrease the effective dimensionality is a further important point.

\section{Conclusion}

In the first part of the paper we discussed the classical linear instantaneous ICA model and
introduced a new algorithm which shows better results than conventional ICA-algorithms. Our 
algorithm should be particularly useful for real world data, since it works with actual 
dependencies between reconstructed sources (as measured by mutual informations), and thus 
easily allows to study the question how independent and unique are the found components. 

In the following we discussed the case where outputs can be grouped together for a meaningful
interpretation. We again saw that MI has some properties which makes it the ideal contrast function,
also for this purpose. 

Finally, when we included time domain structures, we again could use the same estimates of MI, 
with basically the same algorithms. This -- and the excellent results when applied to a fetal 
electrocardiogram -- suggest that our method of basing independent component analysis systematically 
on highly precise estimates of MI is very promising. It is true that our method is slower than 
existing algorithms like FastICA or JADE, but we believe that the improved results justify this 
effort in many situations, in particular in view of the ever-increasing power of digital 
computers.

The software implementation of the MILCA algorithm is freely available online \cite{online}. 

\section*{Appendix}

In this appendix we give two counter examples showing somewhat counterintuitive features
of the MI. In the first example we have two continuous variables, and the joint density 
is constant in an L-shaped domain 
\be
   D = \{[0,l_x]\times [0,\epsilon] \cup [0,\epsilon] \times [0,l_y]\}.
\ee
It is zero outside $D$. It is easily seen that $I(X,Y) \to h$ in the limit $\epsilon 
\to 0$, with $h = p \log p + (1-p) \log (1-p)$ and with $p = l_x/(l_x+l_y)$. In this limit, 
the marginal distributions are superpositions of 
a delta peak at $x$ or $y$ equal to zero, and a uniform distribution on $[0,1]$.
The components have relative weights $l_x:l_y$. The only information about $y$ learned by
fixing $x$ is on which arm the pair $(x,y)$ is located, and for this $h$ bits are sufficient.

On the other hand, any linear transformation applied to the $(x,y)$-plane would give 
an L-shaped figure with at least one oblique arm. For such a distribution knowing 
$x$ would specify $y$ with an accuracy $\sim \epsilon$, and thus $I(X,Y) \sim -\log
\epsilon \to \infty$ for $\epsilon\to 0$. But the covariance between $X$ and $Y$ 
is not zero, hence the minimal MI is reached (for small $\epsilon$) when the correlation
coefficient $r$ is non-zero. A more detailed analysis shows that $I(X,Y)$ of the distribution
rotated by an angle $\phi$ is not symmetric under $\phi\to -\phi$, if $l_x \neq l_y$.

The second example is one of three random variables $X,Y$, and $Z$ which are pairwise 
strictly independent, but globally dependent. For simplicity, the example uses discrete
and indeed binary variables. We have thus 8 probabilities $p(x,y,z)$ for each variable 
being either 0 or 1, and we chose them as $p(0,0,0)=p(1,1,0)=p(0,1,1)=p(1,0,1)=1/8+
\epsilon$ and $p(0,0,1)=p(0,1,0)=p(1,0,0)=p(1,1,1)=1/8-\epsilon$. For this choice
all pairwise probabilities are 1/4, but $I(X,Y,Z) \neq 0$.

Acknowledgements: We thank Drs. Ralph Andrzejak, Thomas Kreuz, and Walter Nadler for numerous 
discussions. H.S. thanks also Andreas Ziehe for invaluable discussions and comments.


\begin{thebibliography}{7}
\bibitem{footnote3} Notice that this value is much lower than the pairwise estimate 
   $\hat{I}(\hat{s}_1,\hat{s}_2) \approx 4$ shown in Fig.~\ref{algocompdep}, which 
   seems to contradict the claim that this MI is dominated by the dependence between
   the first two sources. This is explained by the fact that $\hat{I}(\hat{s}_1,\hat{s}_2)$
   is estimated from much closer neighbors (working in 2 dimensions only), and thus
   is able to resolve much finer details.
\bibitem{footnote5} The matrix elements $A_{ij}$ were uniformly chosen from the interval
   $[0,1]$. Thus they are not normalized and give only relative concentrations,
   but this is irrelevant.
\bibitem{foot7} Although the generalizations to multidimensional ones would be in principle straightforward, 
   we shall in the following discuss only the uniqueness of 1-d output components. In particular,
   we shall treat even those output channels as 1-dimensional which are not independent. This 
   might seem a bit schizophrenic, but it is easier to discuss and nothing is lost in comparison
   with the case where only independent clusters are checked for uniqueness.
\bibitem{footnote} The curves shown in Fig.~\ref{noiseangle} seem to be nearly symmetric around the 
   minimum. But we show in the appendix that this need not be the case in general.
\bibitem{footnote1} We are aware of the fact that the beat rate seen in Fig.~\ref{ECGMILCA} 
   sems much too high, suggesting that the ECG was indeed sampled with $\approx 200$ Hz. But 
   the value of 500 Hz was confirmed by the authors maintaining the URL \cite{ECGdata}. Anyhow, 
   our conclusions are independent of the actual sampling rate.
\bibitem{footnote4} When estimating the individual MIs on the right hand side, one should
   pay attention that the same neighbors are used, i.e. one should not use the same value
   of $k$ in each term (see also footnote \cite{footnote3}).
\end{thebibliography}
\end{document}